\documentclass[prb,twocolumn,showpacs,preprintnumbers,superscriptaddress,amsmath,amssymb,10pt]{revtex4-1}

\usepackage[pdftex]{graphicx}
\usepackage{bm}

\usepackage{mathcomp}
\usepackage{amsmath}
\usepackage{placeins}

\usepackage{capt-of}

\usepackage{gensymb}

\usepackage{multirow}

\usepackage[justification=raggedright,singlelinecheck=false]{caption}

\usepackage{color}

\begin{document}

\title{Interlayer excitons in MoSe$_2$/WSe$_2$ heterostructures from first-principles}
\author{Roland Gillen}\email{roland.gillen@fau.de}
\affiliation{Institute of Solid State Physics, TU Berlin, Hardenbergstr. 36, 10623 Berlin, Germany}
\affiliation{Department of Physics, Friedrich-Alexander University Erlangen-N\"{u}rnberg, Staudtstr. 7, 91058 Erlangen, Germany}
\author{Janina Maultzsch}
\affiliation{Department of Physics, Friedrich-Alexander University Erlangen-N\"{u}rnberg, Staudtstr. 7, 91058 Erlangen, Germany}

\date{\today}

\begin{abstract}
Based on \emph{ab initio} theoretical calculations of the optical spectra of vertical heterostructures of MoSe$_2$ (or MoS$_2$) and WSe$_2$ sheets, 
we reveal two spin-orbit-split Rydberg series of excitonic states below the \textsl{A} excitons of MoSe$_2$ and WSe$_2$ with a significant binding energy on the order of 250\,meV for the first excitons in the series. At the same time, we predict from accurate many-body G$_0$W$_0$ calculations that crystalographically aligned MoSe$_2$/WSe$_2$ heterostructures exhibit an indirect fundamental band gap. Due to the type-II nature of the MoSe$_2$/WSe$_2$ heterostructure, the indirect transition and the exciton Rydberg series corresponding to a direct transition exhibit a distinct interlayer nature with spatial charge separation of the coupled electrons and holes. Our calculations confirm the recent experimental observation of a doublet nature of the long-lived states in photoluminescence spectra of MoX$_2$/WY$_2$ heterostructures and we attribute these two contributions to momentum-direct interlayer excitons at the $K$ point of the hexagonal Brillouin zone and to momentum-indirect excitons at the indirect fundamental band gap. Our calculations further suggest a noticeable effect of stacking order on the {electronic band gaps and on the peak energies of the interlayer excitons and their oscillation strengths.} 
\end{abstract}

\pacs{78.20.Bh; 73.22.-f; 71.15.Mb; 71.35.-y; 11.10.St}
\maketitle

\section{Introduction}
Transition metal dichalcogenides (TMDC) of molybdenum and tungsten are promising members of the family of layered materials due to the versatility of their physical properties. On one hand, they are intrinsic semiconductors in bulk and few-layer phases, with a direct fundamental band gap in the monolayer form. This band gap transition, accompanied by strong excitonic effects, leads to an enhancement of photoluminescence quantum yield for decreasing material thickness~\cite{mak-2010,splendiani-2010,MoSe2WSe2MoS2-PL,WS2-PL,MoS2-tunable-PL,scheuschner-2014}. It inspired applications of TMDCs in novel thin and flexible optoelectronic devices, such as photodiodes~\cite{led1,led2}, photodetectors~\cite{photodetector1} or single-photon emitters~\cite{single-photon2,single-photon3,single-photon4,single-photon5}.
On the other hand, the heavy transition-metal atoms possess a significant spin-orbit interaction that causes a split of the valence band edge. The associated coupled spin-and valley physics open a path towards a combination of spin- and valleytronics~\cite{spinvalley-1, spinvalley-2}. 

An additional advantage of layered materials such as TMDCs is the saturated covalent bonds within one layer and non-covalent binding between the layers, which allows for atomically sharp and stress-free interfaces between two different layered materials, \emph{e.g.} in p-n junctions made of MoSe$_2$ and WSe$_2$~\cite{WSe2MoSe2-1} or similar materials. 
Heterostructures of transition metal dichalchogenides with other materials thus offer a powerful path to engineer flexible compound materials and devices with desired optical and electronic properties. The vertical or lateral combination of sheets of Mo and W TMDCs recently gained particular interest due to the observation of long-lived excitonic states in the photoluminescence spectra of MoS$_2$/WS$_2$~\cite{MoS2WS2-1}, MoS$_2$/WSe$_2$~\cite{MoS2-WSe2-1}, MoSe$_2$/WSe$_2$~\cite{WSe2MoSe2-2,WSe2MoSe2-3,nayak-MoSe2WSe2,miller-indirectexc} and MoSe$_2$/WS$_2$~\cite{MoSe2WS2-1} heterostructures. Due to the expected type-II alignment~\cite{kang-offsets} of MoX$_2$ and WX$_2$ bands (with X=S, Se, Te), this observation has been attributed to interlayer excitons with a spatial separation of the coupled electrons and holes, which is of interest for applications in photovoltaics.

Despite this interest, to the best of our knowledge, theoretical confirmations of this interpretation from \emph{ab initio} are scarce and indirect so far~\cite{Komsa-2013,palumno-2015,spataru-interlayercoupling}. 
Another recent study~\cite{lantini-2017} employed a generalized Mott-Wannier model to vertical MoS$_2$/BN/WSe$_2$ heterostructures, which is limited to the exciton binding energy.
This lack of studies in the literature can be understood from the complications of \emph{ab initio} simulations of the excitonic spectra in 2D TMDC heterostructures due the lowered symmetry, the necessity of careful treatment of the quasi-two-dimensional dielectric screening and the importance of strong spin-orbit coupling for both the band alignment in stacked heterostructures and the nature of excitonic transitions. 

We thus report fully \emph{ab initio} theoretical simulations of the optical spectra of MoSe$_2$/WSe$_2$ and MoS$_2$/WSe$_2$ heterostructures under full inclusion of electron-hole and spin-orbit interaction. Our computations allow a direct estimate of the expected exciton binding energy and give access to the excitonic wave functions. By considering three distinct stacking orders we assess the possible influence of arbitrary alignment of the layers in experimental conditions. 

\section{Method}
\begin{figure*}
\centering
\includegraphics*[width=0.9\textwidth]{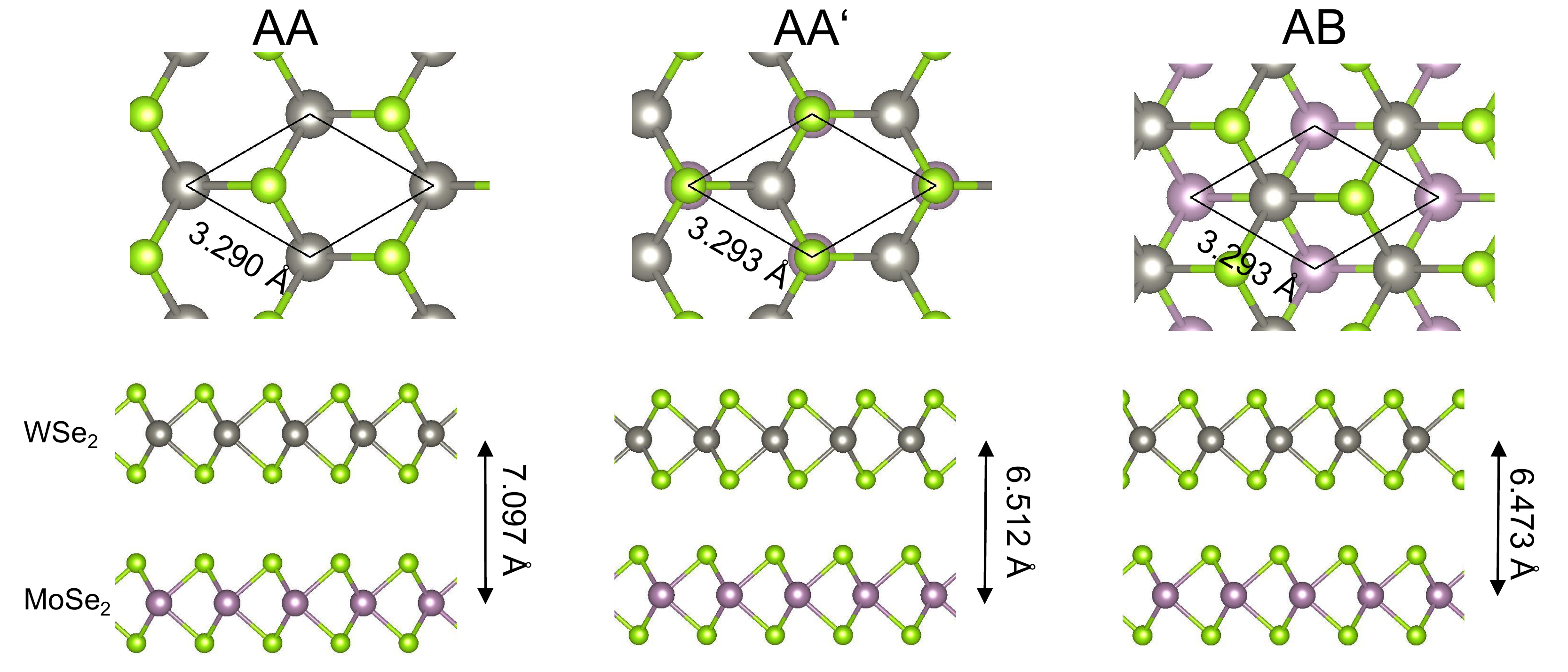}
\caption{\label{fig:MoSe2-WSe2-structures} (Color online) Geometries of the computed heterostructures of monolayers of MoSe$_2$ and WSe$_2$ with AA, AA' and AB stackings. Indicated are the optmimized in-plane lattice constants and layer distances. The typical stacking in bilayer and bulk MoS$_2$, MoSe$_2$ and WSe$_2$ corresponds to AA'.}
\end{figure*}
We calculated the groundstate electronic wavefunctions and bandstructures with the Quantum Espresso package\cite{qe} in the Perdew-Becke-Ernzerhof (PBE) approximation, using fully relativistic normconserving pseudopotentials including the semi-core $s$ and $p$ orbitals of Mo and W. Atomic positions and cell parameters were relaxed with inclusion of semi-empirical van-der-Waals corrections from the PBE+D3\cite{d3-2} scheme.

The dielectric functions including electron-hole interactions and spin-orbit interaction were computed by solving the Bethe-Salpeter Equation (BSE) using the YAMBO code\cite{yambo} on a discrete grid of 21x21x1 k-points. 20 valence bands and 20 conduction bands were included for the calculation of the absorption spectra. The static dielectric function was calculated with 600 empty bands and a cutoff for the response function of 200\,eV. The electronic bandstructures from density functional theory (DFT) were corrected by G$_0$W$_0$ quasiparticle energies using 1500 unoccupied bands. The exchange and correlation contributions where extrapolated to infinite cutoff energy. In both GW and BSE calculations, we used the effective energy technique~\citep{bg-eet} to include contributions from high-energy unoccupied bands.

We found it crucial for quasi-two-dimensional materials to properly treat the singularity of the Coulomb interaction in order to obtain interpretable quantitative results. We truncated the Coulomb interaction in the non-periodic direction following the method from Ref.~\onlinecite{ismail-beigi} and averaged the head of the screened Coulomb interaction $W$, \emph{i.e.} the contribution $W(\mathbf{q}\rightarrow\mathbf{0},\mathbf{G}\rightarrow\mathbf{0},\mathbf{G}'\rightarrow\mathbf{0})$ by using a model function for the dielectric screening in the vicinity of the $\Gamma$ point.
Our results suggest that different treatments of the Coulomb interaction and the dielectric screening close to the $\Gamma$ point accounts for some of the spread in calculated excitonic binding energies in TMDCs found in the literature. We refer to the supplementary information for further computational details.

\section{Results and Discussion}
\begin{figure*}
\centering
\includegraphics*[width=0.32\textwidth]{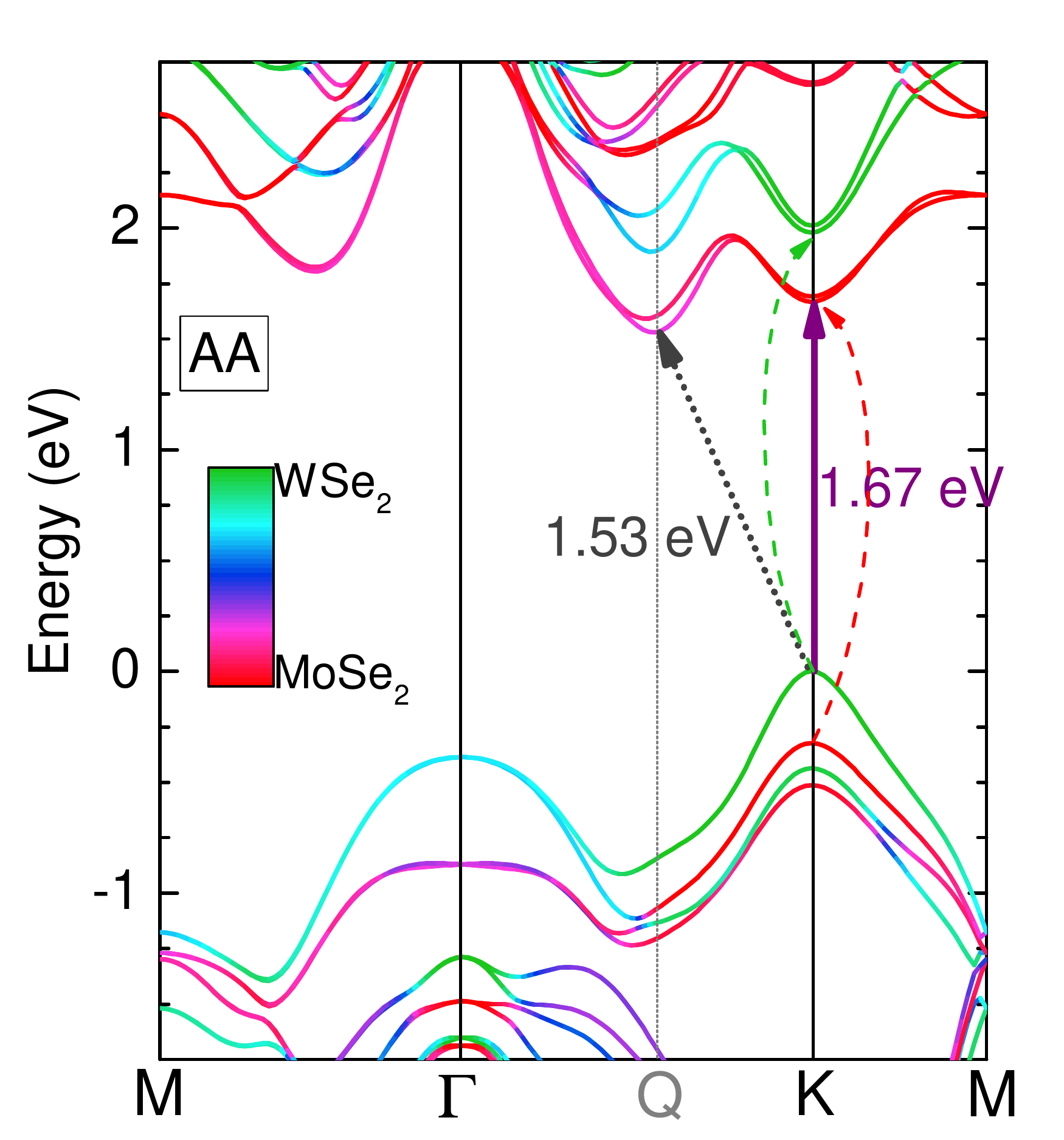}
\includegraphics*[width=0.32\textwidth]{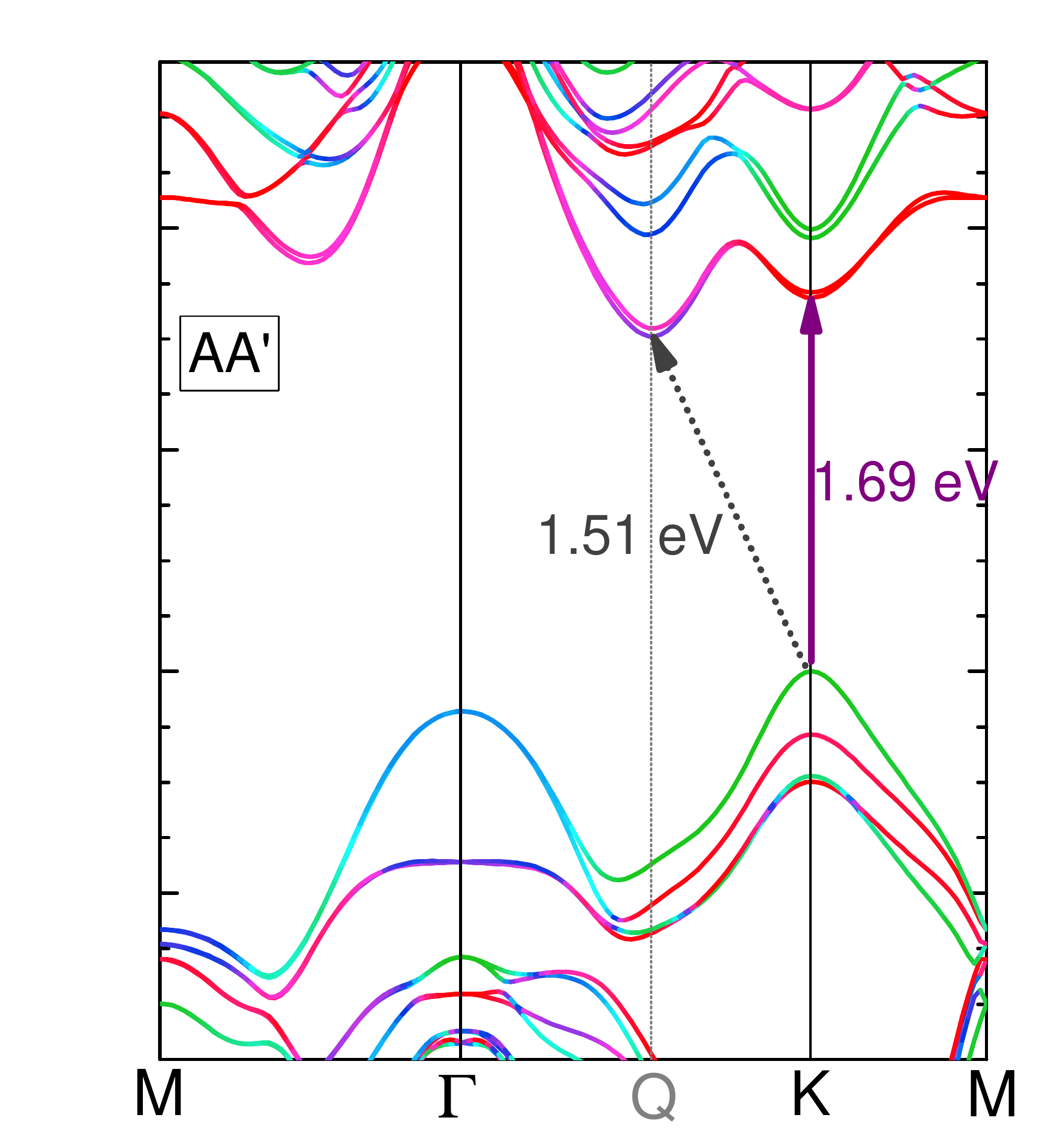}
\includegraphics*[width=0.32\textwidth]{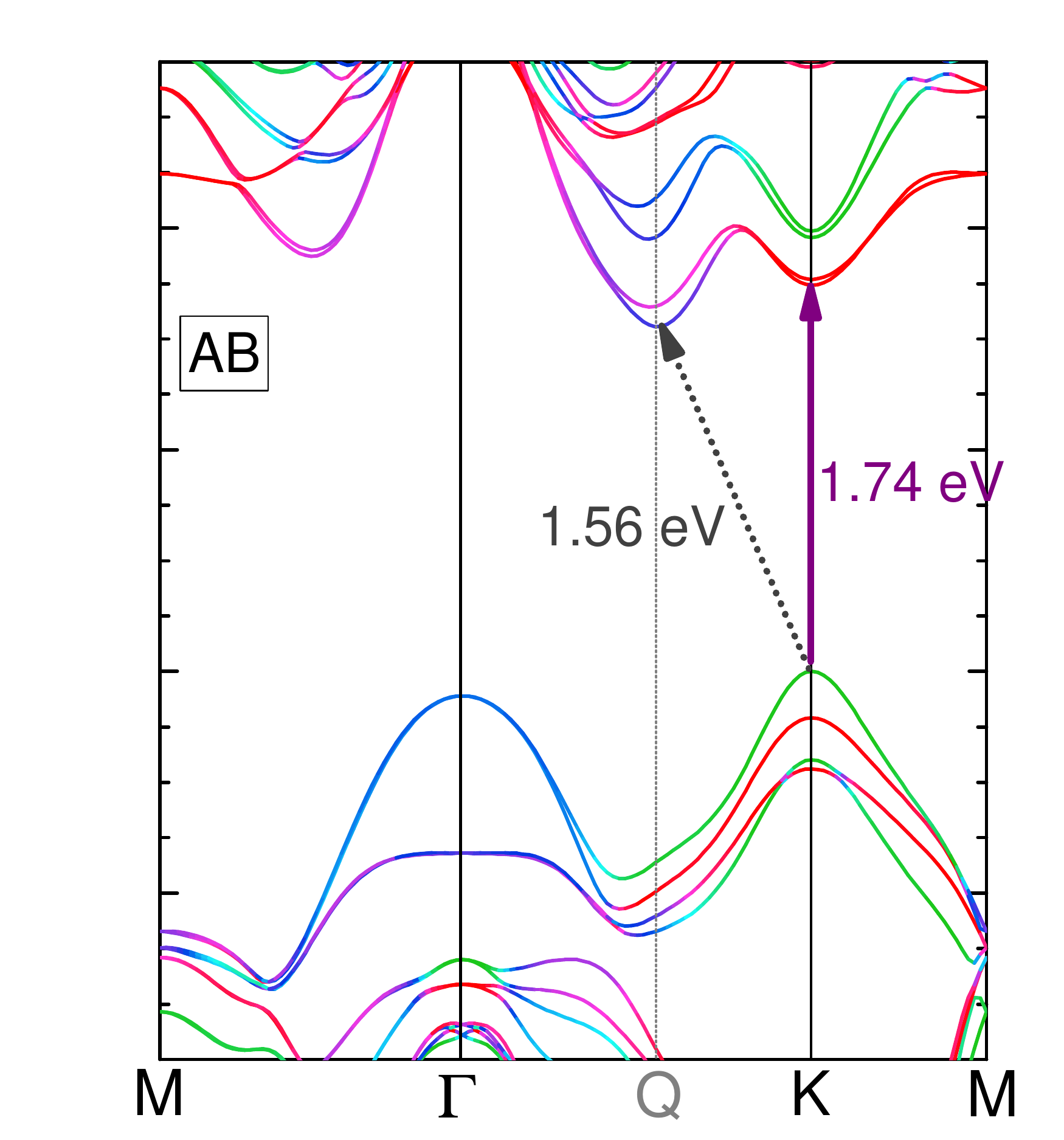}
\caption{\label{fig:MoSe2-WSe2-bands} (Color online) Electronic bandstructures for three different stackings of monolayers of MoSe$_2$ and WSe$_2$. All computations were corrected by G$_0$W$_0$ quasiparticle energies and by full inclusion of spin-orbit interactions. The zero-of-energy is set to the valence band maximum for each stacking. The color scale depicts the relative contributions of the materials to the bands.}
\end{figure*}
\begin{figure}[b]
\centering
\includegraphics*[width=0.99\columnwidth]{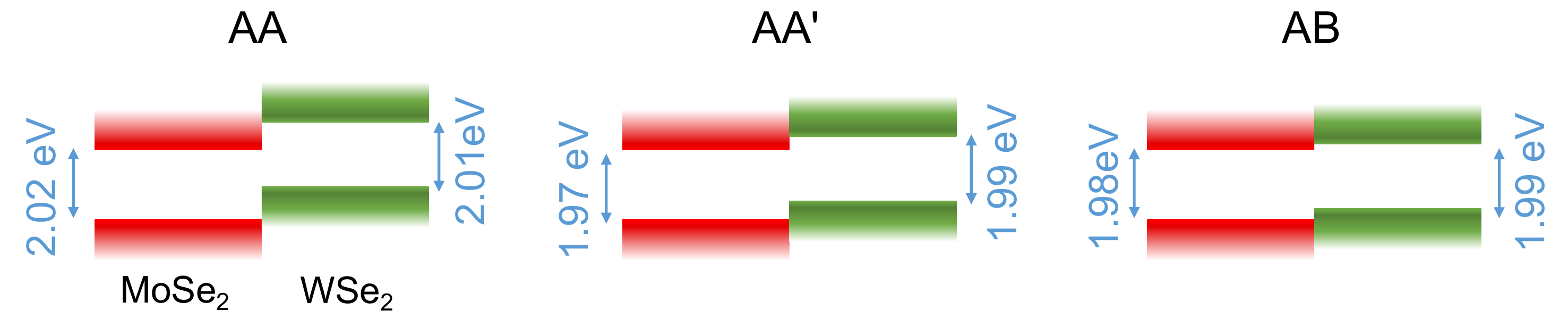}
\caption{\label{fig:MoSe2-WSe2-alignment} (Color online) Schematic band alignment for the three studied stacking orders with intralayer band gaps at the $K$ point. The band gap values energies given are taken from the calculations shown in Fig.~\ref{fig:MoSe2-WSe2-bands}.}
\end{figure}
\begin{table}[t]
\caption{\label{tab:bandoffsets} Band offsets obtained from G$_0$W$_0$ calculations. For the conductions band, we give the band offsets between MoSe$_2$ and WSe$_2$ dominated bands at the $Q$ and the $K$ points.}
\begin{ruledtabular}
\begin{tabular}{c c c c}
Band offsets$\backslash$ Stacking order & AA & AA' & AB\\
\hline
$\Delta E_V$ (meV)& 326 & 284 & 209\\ 
$\Delta E_C^K$ (meV)& 313 & 270 & 216\\
$\Delta E_C^Q$ (meV)& 362 & 458 & 394\\
\end{tabular}
\end{ruledtabular}
\end{table}

Experimental samples of vertical heterostructures are often fabricated from exfoliation procedures, where individual layers are stacked manually, typically exhibiting an arbitrary stacking order.
The local layer alignment can affect the electronic bandstructure of the composite material through the formation of interface dipoles (adding a relative shift of the bandstructures of the two materials) and interlayer hybridization of orbitals. The latter should be particularly strong at $\Gamma$ and similar points in the Brillouin zone that possess considerable contributions from chalcogen \textit{p} states, as has been shown for mixed heterostructures~\cite{interlayer-coupling1} and twisted bilayers of the same material~\cite{interlayer-coupling2}.

In order to obtain an estimate of the effects of relative alignment on the electronic band structures and simulated optical spectra of MoSe$_2$-WSe$_2$, we consider three different stacking orders that all have the advantage of preserving the hexagonal symmetry of the pure materials: The AA stacking corresponds to a zero degree rotation of the WSe$_2$ with respect to the MoSe$_2$ layer. AA' is the most stable stacking in MoSe$_2$ and WSe$_2$ bilayer and bulk with a 180$\degree$ relative rotation. The AB stacking is an AA stacking with a relative shift between MoSe$_2$ and WSe$_2$ layers by $a/\sqrt{3}$, where $a$ is the in-plane lattice constant, and the preferred stacking of rhombohedral (3R-) MoS$_2$, graphite and hexagonal boron nitride.

Figure~\ref{fig:MoSe2-WSe2-structures} shows the optimized geometries obtained from our PBE-D3 calculations. The different stackings have only a minor effect on the in-plane lattice constants, with a variation of about 0.1\%. The obtained lattice constants of 3.290-3.293\AA\space are only slightly changed compared to the monolayer materials and in good agreement with the experimental lattice constants of bulk MoSe$_2$ and WSe$_2$ of 3.28-3.29\AA~\cite{MoSe2-WSe2-latconsts}. On the other hand, the interlayer distance shows a significantly stronger variation. For AA stacking, the chalcogen atoms of the two layers are right on top of each other, which leads to an increased interlayer distance of about 7.1\,\AA. For the other stackings, the chalcogen atoms are aligned with the metal atoms of the neighbouring layer (AA') or are partially aligned with the centers of the neighbouring metal-selenium hexagons. The resulting interlayer distances of 6.51\AA~(AA')\space and 6.47\AA~(AB) are only slightly larger than the interlayer distances in AA' stacked bilayer MoSe$_2$ and WSe$_2$. 
We used these optimized geometries to calculate the electronic bandstructures of the three heterostructures, shown in Fig.~\ref{fig:MoSe2-WSe2-bands}. Full spin-orbit-interaction was included to correctly describe the large spin-orbit splitting of valence and conduction bands at the $K$-point in the hexagonal Brillouin zone. 
The bands at the $K$ point show only negligible signs of hybridization for the three stacking orders, due to the dominant contributions from Mo and W $d$ states. In accordance to previous studies~\cite{kang-offsets}, the bands of the MoSe$_2$ and WSe$_2$ are shifted relative to each other such that the global valence band maximum is in the WSe$_2$ layer (green color in Fig.~\ref{fig:MoSe2-WSe2-bands}) and the conduction band valley is in the MoSe$_2$ layer (shown in red). The stacking order affects the band alignment and the size of the direct interlayer band gap at $K$, as shown in Fig.~\ref{fig:MoSe2-WSe2-alignment}. We compiled the band offsets for the three stacking orders in Tab.~\ref{tab:bandoffsets}.
 
On the other hand, the main qualitative differences of the electronic band structures due to different layer stackings are found at the valence band edge at the $\Gamma$ point and the conduction band valley around the half-way point ($Q$) in the $ \Gamma$-$K$ direction. At the $\Gamma$ point, the bands of the two materials are prone to hybridization due to contributions from Se $p$ states, causing split bands of mixed MoSe$_2$ and WSe$_2$ character. Interestingly, the interlayer interaction in all three stacking orders appears to be strong enough to pull down the conduction band valley at the $Q$ point, similar to the case of homobilayers of MoSe$_2$ and WSe$_2$. This shifts the global conduction band minimum from the $K$ point to the $Q$ point. The second lowest conduction band at the $Q$ valley is dominated by Mo $d_{\mathrm{x^2-y^2}}$ and Se $p_{\mathrm{x}}$/$p_{\mathrm{y}}$ states of the MoSe$_2$ layer, while the lowest conduction band obtains larger contributions from the WSe$_2$ layer for decreasing interlayer distance but mainly remains within the MoSe$_2$ layer.
Our simulations thus suggest that MoSe$_2$ and WSe$_2$, especially for random stacking configurations, form a type-II heterostructure with an \textit{indirect} fundamental band gap~\cite{Note1}.
This introduces the possibility of momentum-indirect excitons with strong interlayer nature as candidates for the experimentally observed interlayer excitons.
\begin{figure}
\centering
\includegraphics*[width=0.95\columnwidth]{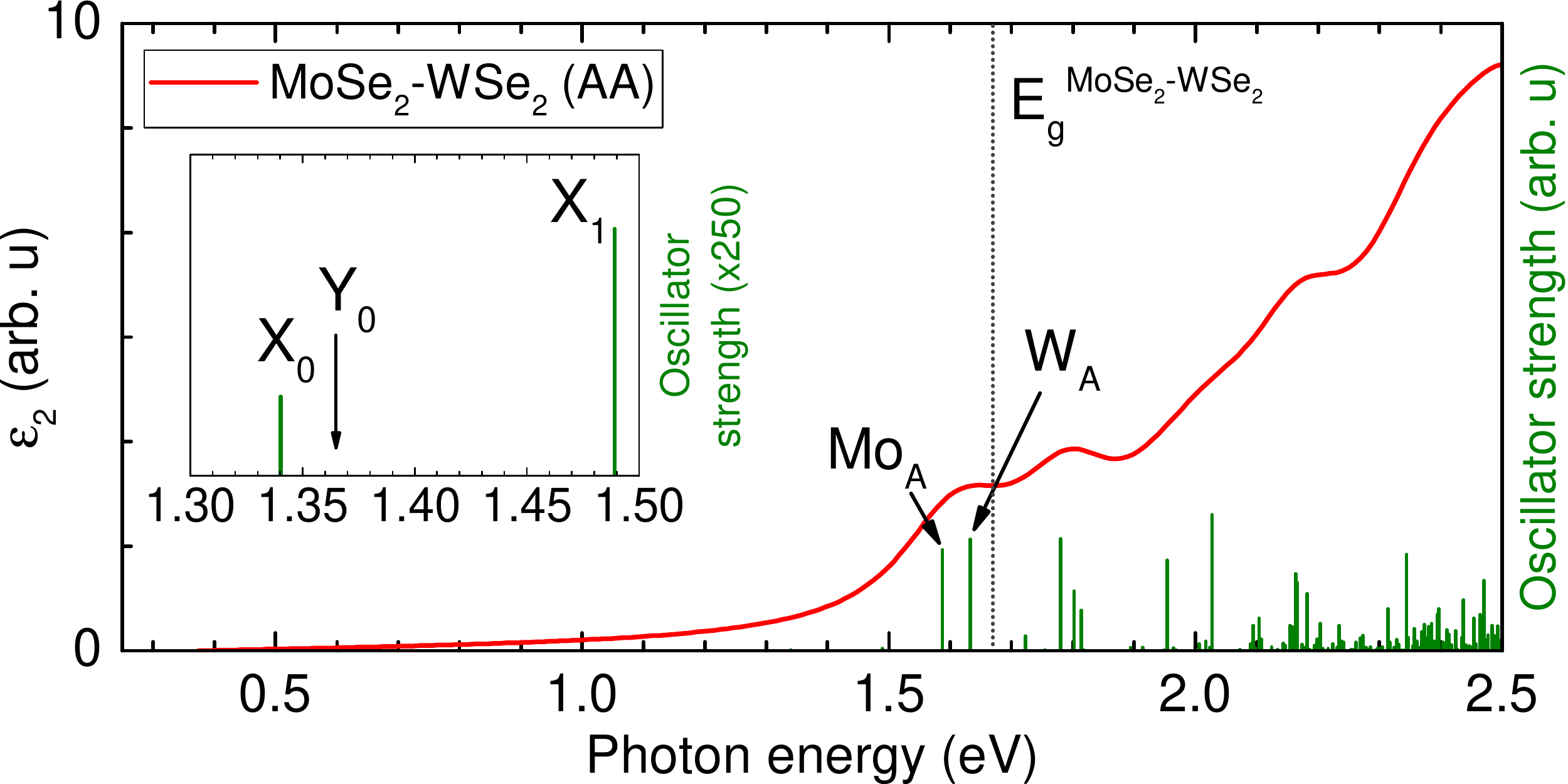}
\includegraphics*[width=0.95\columnwidth]{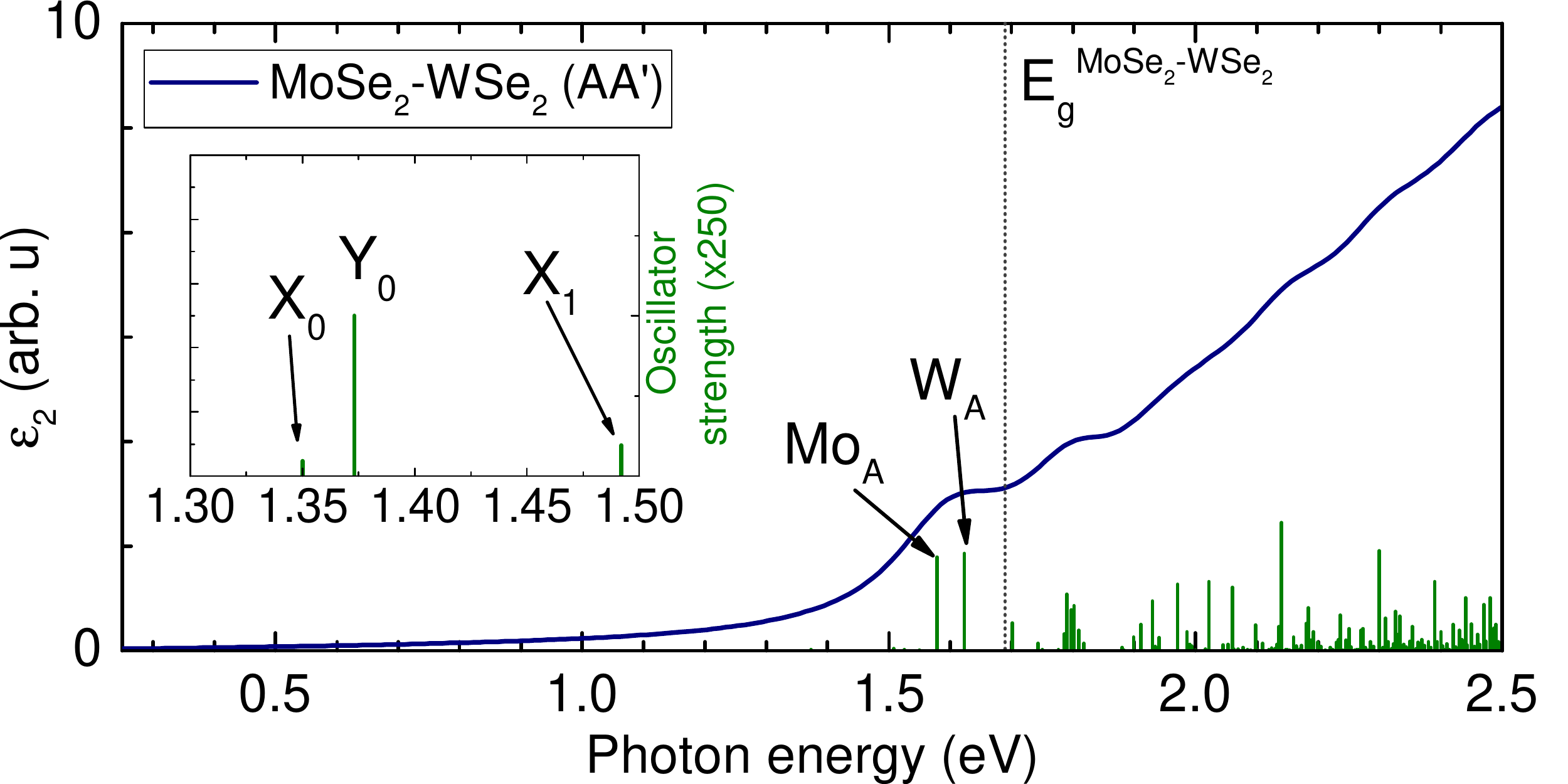}
\includegraphics*[width=0.95\columnwidth]{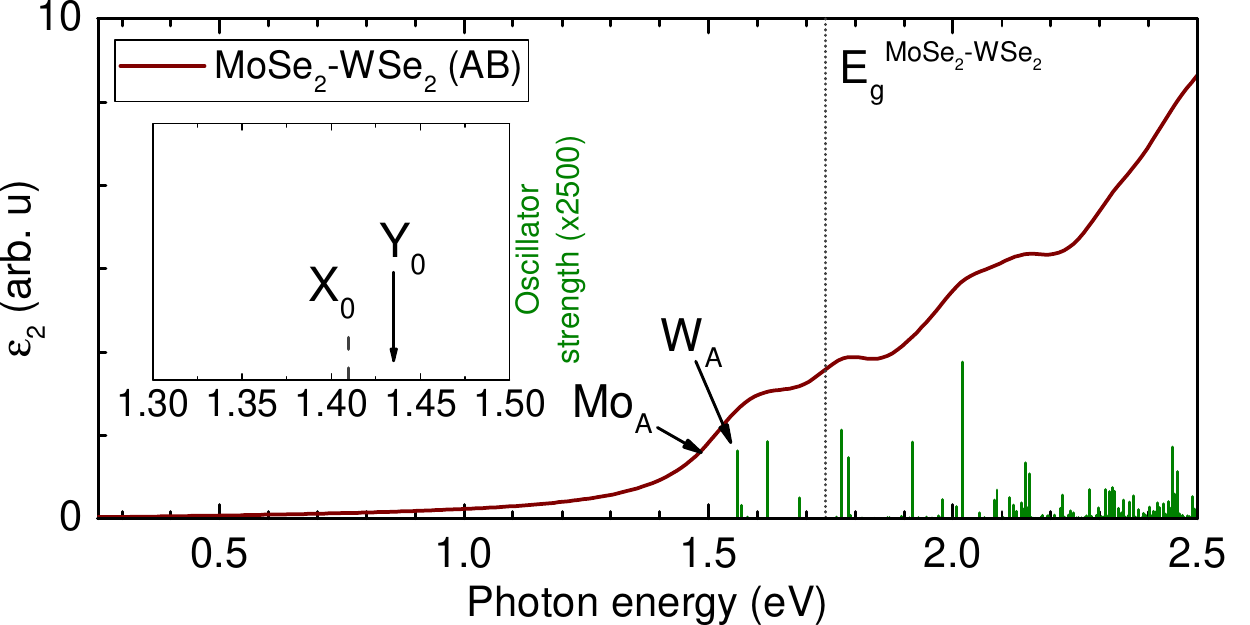}
\caption{\label{fig:MoSe2-WSe2-absorption} (Color online) Calculated dielectric functions $\epsilon_2$ for three stackings with electron-hole effects and spin-orbit coupling and light polarization \textit{parallel} to the layer plane of the heterostructure. Green bars show the optical oscillator strenghts of the constituting excitonic and band transitions. The insets show the energy region around the zero-order interlayer excitonic peaks. The dashed line in the inset for AB stacking shows the \textsl{X}$_0$ transition for light polarization \textit{perpendicular} to the surface. The intensity for light polarized parallel to the plane vanishes. Note the different scale in the inset for AB stacking.}
\end{figure}
\begin{figure*}
\centering
\includegraphics*[width=0.95\textwidth]{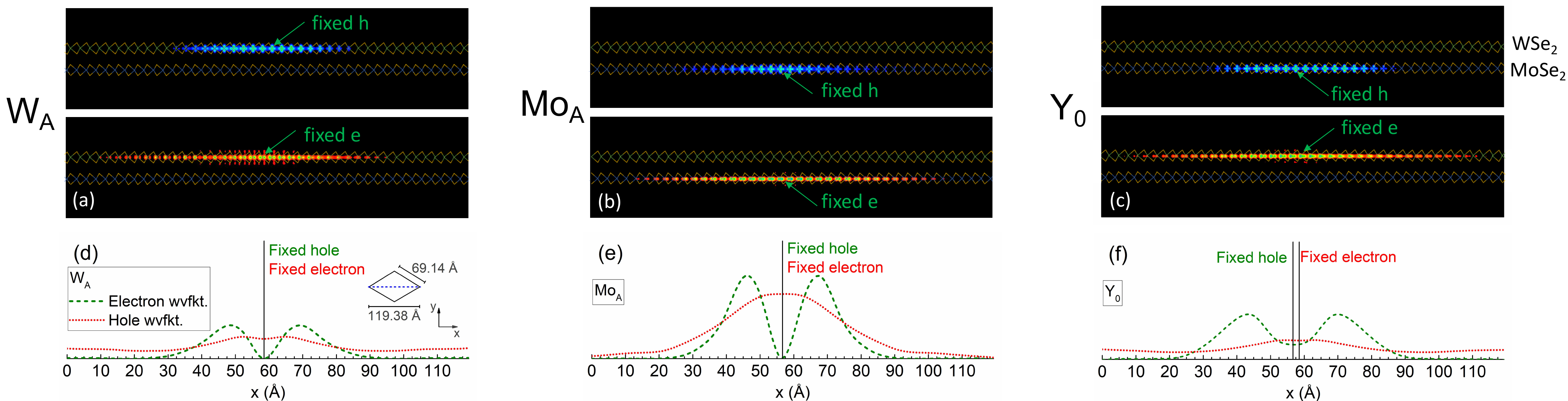}
\caption{\label{fig:MoSe2-WSe2-wavefunctions} (Color online) Electron (blue) and hole (red) contributions to the excitonic wave functions of the (a) \textsl{W$_A$}, (b) \textsl{Mo$_A$} and (c) \textsl{X$_0$} excitons of an AA'-stacked MoSe$_2$/WSe$_2$ heterostructure. The atomic structure of the two layers is indicated, in each panel, the upper layer is WSe$_2$ and the lower layer is MoSe$_2$. The excitonic wave functions were computed for a supercell of 21x21 unit cells and projected onto the x-z plane. For the electron (e) [hole (h)] contributions, the hole [electron] was fixed at the Mo [W] atom indicated by the green arrows. (d)-(f) Envelope functions of the electron and hole parts of the excitonic wavefunctions from (a)-(c) along a path in x-direction depicted as dashed line in the inset in (d). The path was chosen to contain the fixed electron/hole. The shape of the electronic part of the excitonic wavefunctions agrees with Ref.~\onlinecite{wirtz-mos2-excitons}. Top views of the excitonic wavefunctions and a discussion of the spatial extension of the electron and hole parts can be found in the supplementary material.}
\end{figure*}
\begin{table}[t]
\caption{\label{tab:excitations} Peak positions ($E$) and binding energies ($E_b$) of selected excitonic transitions with respect to the corresponding electronic band gaps ($E_g$) for AA' stacking. For the binding energy, we give two different values for calculations with ($E_b^{\text{SOI}}$) and without ($E_b^{\text{no SOI}}$) spin-orbit coupling.  The \textsl{X$_0$} transition occurs between the highest valence band (in the WSe$_2$ layer) and the lowest conduction band (in the MoSe$_2$ layer) at the $K$ point, while the \textsl{Y$_0$} transition occurs between the highest valence and the second lowest conduction band. }
\begin{ruledtabular}
\begin{tabular}{c c c c c}
Excitation & $E_g$ (eV) & $E$ (eV) & $E_b^{\text{SOI}}$\footnote{Obtained with a 21x21 k-point sampling. Due to the spatial extent of the excitonic wavefunctions, the binding energies are overestimated for this sampling} 
(eV) & $E_b^{\text{no SOI}}$\footnote{Obtained with a 33x33x1 k-point sampling that yields accurate exciton binding energies. Inclusion of spin-orbit coupling has negligible effects on the results.} (eV)\\
           &        & \multicolumn{2}{c}{21x21 k-point grid} & 33x33 grid\\
\hline
Mo$_A$ & \multirow{3}{*}{1.978} & 1.578 & 0.41 & \textbf{0.307}\\
Mo$_A^*$ &  & 1.701 & 0.277 & \textbf{0.176}\\
Mo$_A^{**}$ &  & 1.733 & 0.245 & \textbf{0.131}\\
W$_A$ & \multirow{3}{*}{1.994} & 1.623 & 0.377 & \textbf{0.267}\\
W$_A^*$ &  & 1.791 & 0.203 & \textbf{0.156}\\
W$_A^{**}$ &  & 1.835 & 0.159 & \textbf{0.121}\\
\hline
Mo$_B$ & 2.22 & 1.80 & 0.42 & -\\
W$_B$ & 2.432 & 2.06 & 0.372 & -\\
\hline
X$_0$ & 1.685 & 1.350 & 0.335 & \textbf{0.251} \\
X$_1$ &  ($K\rightarrow K$) & 1.492 & 0.193 & \textbf{0.148}\\
Y$_0$ & 1.709 & 1.373 & 0.336 & - \\
Y$_1$ & ($K\rightarrow K$) & 1.509 & 0.200 & -\\
\hline
indirect & 1.51 & &\\
bandgap & ($K\rightarrow Q$) &&&
\end{tabular}
\end{ruledtabular}
\end{table}
\begin{figure}[b]
\centering
\includegraphics*[width=0.95\columnwidth]{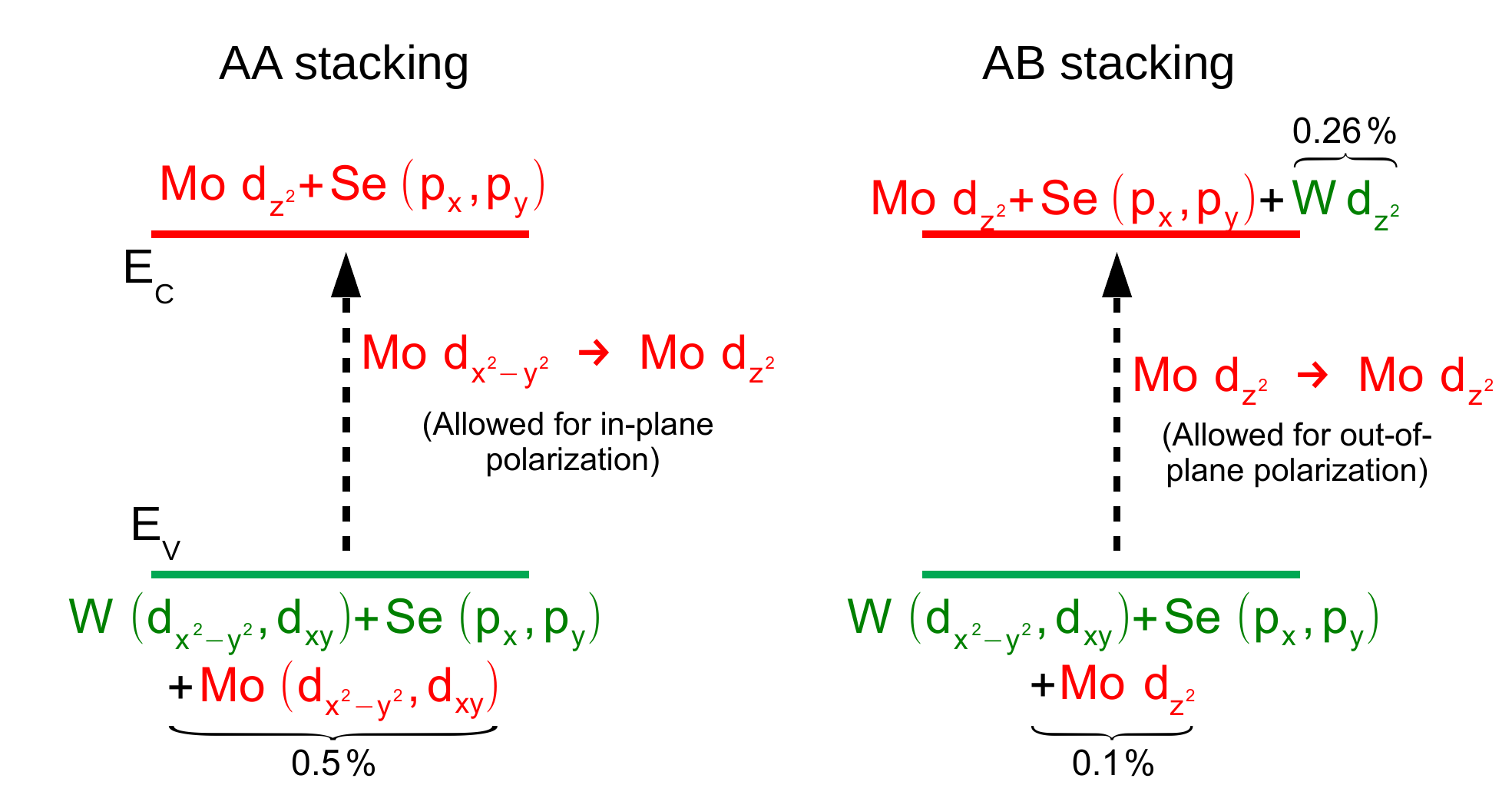}
\caption{\label{fig:MoSe2-WSe2-contributions} (Color online)Schematic composition of the band edges $E_V$ and $E_C$ at the $K$ point for AA and AB stacking. Small stacking-dependent mixing of states of the neighboring layer occurs due to weak interlayer hybridization. For each stacking order, the main contribution to the optical oscillator strength of interlayer transitions is shown by a dashed arrow.}
\end{figure}

We will now show that, in addition to the indirect transition discussed above, a second interlayer transition at the $K$ and $K'$ points leads to strong excitonic effects and can be attributed to the experimentally observed interlayer transition. 
Based on the electronic structures in Fig.~\ref{fig:MoSe2-WSe2-bands}, we used the excitonic Bethe-Salpeter equation (BSE) to compute the corresponding absorption spectra from direct band transitions~\cite{Note2}
including electron-hole effects for the three stacking orders. An advantage of this theoretical approach is the ability to decompose the obtained spectra into the contributing band transitions. As shown in Fig.~\ref{fig:MoSe2-WSe2-absorption}, for AA, AA' and AB stacking, the absorption onset is dominated by a number of excitonic contributions with high oscillator strengths. The lowest-energy bright contribution, \textsl{W$_A$}, originates from a transition between the valence band top and the fourth conduction band at the \textit{K} point and corresponds to the \textsl{A} exciton in isolated single-layer WSe$_2$. In Fig.~\ref{fig:MoSe2-WSe2-wavefunctions}~(a) we show the excitonic wavefunction of \textsl{W$_A$}. Both the electron and hole parts of the excitonic wave function are localized within the WSe$_2$ layer. 
Similarly, the contribution \textsl{Mo$_A$} corresponds to the \textsl{A} exciton of single-layer MoSe$_2$ and appears at a slightly higher energy due to the difference in spin-orbit splitting of the valence band top for MoSe$_2$ and WSe$_2$. The excitonic wave function of \textsl{Mo$_A$}, Fig.~\ref{fig:MoSe2-WSe2-wavefunctions}~(b), is localized within the MoSe$_2$ layer.

The computational expense of BSE calculations including spin-orbit coupling limited us to a grid of 21x21 k-points for the Brillouin zone integration, which might be insufficient to yield fully converged values for the exciton binding energies. As the spin-orbit interaction only affects the absolute energy of the \textsl{A} excitons but not the binding energy, we hence performed additional calculations without spin-orbit coupling but with denser 33x33 k-point grids. For AA' stacking, we derive exciton binding energies of 267\,meV and 307\,meV for \textsl{W$_A$} and \textsl{Mo$_A$}, respectively, with similar results for the AA and AB stacking orders.
The exciton binding energies are reduced as compared to the isolated single-layer materials (0.48\,eV for MoSe$_2$~\cite{exciton-paper}, 0.46\,eV for WSe$_2$), due to dielectric screening from the neighboring layer in the heterostructure. We will now show that these transitions can be identified with the \textit{inter}layer excitons as proposed from experiments.

In addition to the excitonic contributions from \textit{intra}layer transitions discussed above, our calculations reveal further contributions with low oscillator strenghts at energies below the \textsl{A} excitons. These contributions form two Rydberg-like series of electronic transitions at the fundamental band gap (labeled \textsl{X$_n$} in the insets to Fig.~\ref{fig:MoSe2-WSe2-absorption}) and between the valence band maximum and the second conduction band (\textsl{Y$_n$}). 
The energy of the 'ground state' (i.e. \textit{n}=0) contributions is well below the energy of the fundamental band transition, indicating an excitonic state, and is relatively independent of the stacking order in the heterostructure. 
As the involved valence band maximum and conduction band minimum are composed of WSe$_2$ and MoSe$_2$ states, respectively, we attribute $X_n$ and $Y_n$ to \textit{inter}layer excitons with a distinct charge separation. This charge separation is clearly seen in the plot of the exciton wave function of \textsl{Y$_0$} (for AA' stacking) in Fig.~\ref{fig:MoSe2-WSe2-wavefunctions}~(c): the electronic part of the exciton wave function is confined to the MoSe$_2$ layer, while the hole contribution is confined to the WSe$_2$ layer. We note that due to the 180$\degree$ relative rotation of the layers in the AA' stacking, the $K$ point of the MoSe$_2$ layer is rotated onto the $K'$ point of the WSe$_2$ layer (and vice versa). Due to the swapped order of the spin-orbit split bands at $K'$ compared to the $K$ point, the $Y_0$ transition in AA' stacking hence corresponds to the $X_0$ peak for the AA and AB stackings.

The difference in fundamental band gaps induces small relative shifts of the transition energies of X$_0$ and Y$_0$ for the three different stackings, see Fig.~\ref{fig:MoSe2-WSe2-absorption}. Using a denser grid of 33x33 k-points and neglecting spin-orbit interaction, we estimate the binding energy of \textsl{X$_0$} and \textsl{Y$_0$} to be on the order of 250\,meV for the three different stacking orders, which is of a similar magnitude as the value of 280\,meV derived from a Mott-Wannier model for a vertical MoS$_2$/WSe$_2$ heterostructure~\cite{lantini-2017}. The different stacking orders induce a small variation of 10\,meV between the highest (AB stacking) and lowest (AA stacking) binding energy.The obtained peak energies and the energies compared to the corresponding electronic band gaps are summarized for the AA' stacking order in Tab.~\ref{tab:excitations}.
Surprisingly, the binding energies of \textsl{X$_0$} and \textsl{Y$_0$} exciton are very similar to the binding energies of the intralayer excitons as discussed above. This appears counterintuitive due to the spatial separation of electrons and holes for the interlayer excitons. On the other hand, the spatial extension of the excitonic wavefunctions within the layers is on the order of several nm and hence significantly larger than the distance of the two layers, and the dielectric screening in the heterostructure is anisotropic. It is thus possible that the influence of the spatial separation on the electron-hole interaction is compensated by a reduction of the Coulomb screening in the interstitial region between the MoSe$_2$ and WSe$_2$ layers.

While the peak and exciton binding energies of the spatially indirect (and momentum direct) excitons show relatively weak dependence on the stacking order, the oscillator strengths vary quite significantly. In particular, they almost completely vanish for AB stacking in case of light polarized parallel to the plane of the 2D heterostructure. Interestingly, the \textsl{X}$_0$ transition is 'activated' for light polarized \textit{perpendicular} to the surface, refer to the inset of Fig.~\ref{fig:MoSe2-WSe2-absorption}~(c). Similarly, a number of nominally spin-forbidden transitions gain oscillation strength under these conditions. This motivates a more detailed study of the optical selection rules and spinorial symmetries for different stacking orders. The found strong dependence of oscillator strength on stacking order might be one of the reasons for the experimental perception that interlayer excitons cannot be always detected. 

In order to better understand the origin of these observations, we projected the matrix elements for interlayer transitions at $K$ (neglecting electron-hole interaction effects) to a basis of atomic orbitals. Based on our calculations, it is reasonable to assume that the different polarization behaviors arise from weak interlayer hybridization at the $K$ point that causes a small, stacking dependent, mixing of Mo $d$ states into the valence band maximum, see Fig.~\ref{fig:MoSe2-WSe2-contributions}. Our calculations suggest that the optical matrix elements are dominated by transitions between the spilled-over Mo $d$ states with the conduction band minimum. This would explain generally observed much lower oscillation strengths of the \textsl{X$_0$} and \textsl{Y$_0$} compared to the intralayer \textsl{Mo$_A$} and \textsl{W$_A$} transitions, which is in agreement with recent photocurrent measurements on MoSe$_2$/WSe$_2$ heterostructures~\cite{WSe2MoSe2-3}, and should apply similarly for spatially indirect transitions at the indirect fundamental band gap of the system. It would also explain why the polarization dependence of $X_0$ is different from that of \textsl{Mo$_A$} and \textsl{W$_A$} for AB stacking. We found a similar Rydberg series of interlayer excitonic states with low oscillation strengths for a MoS$_2$/WSe$_2$ bilayer heterostructure as well. We refer to the supplementary material for calculated absorption spectra for light polarization perpendicular to the heterostructure slab, details about the projection of the optical matrix elements and for the calculations on the MoS$_2$/WSe$_2$ bilayer heterostructure.

The very small oscillator strengths suggest that the interlayer excitons are not visible in absorption measurements. On the other hand, they might be visible as excitonic states in photoluminescence (PL) measurements due to radiative recombination of electron-hole pairs that have relaxed to the band extrema. In this context, our calculations suggest two possible contributions to the experimentally observed interlayer exciton: {(i) Spatially indirect, momentum direct recombinations of the \textsl{X$_0$} and \textsl{Y$_0$} excitons at the $K$ point with a substantial binding energy of 250\,meV. (ii) A spatially indirect, momentum indirect recombination at the indirect fundamental band gap. Based on a comparison of our calculated indirect band gap of 1.5\,eV with the position of the interlayer PL peak, we estimate the exciton binding energy to be of order 0.1-0.2\,eV. This estimate appears reasonable in light of the significant binding energy of the momentum direct interlayer excitons at the $K$ point and is further supported by the good prediction of the electronic band gaps and peak positions of the intralayer \textsl{A} excitons compared to experiments.}
The first contribution might be enabled by inefficient relaxation of excited electrons from the $K$ to the $Q$ conduction band valley; similar to the observations from photoluminescence experiments on homobilayers of MoSe$_2$ and WSe$_2$, where the MoSe$_2$ and WSe$_2$ \textsl{A} peak appears together with the indirect transition. On the other hand, from a naive point of view, the efficiency of the first pathway should be limited by its two-fold indirect nature.Our results are in good qualitative and quantitative agreement with the recent photoluminescence experiments by Miller \emph{et al.}~\cite{miller-indirectexc}, who showed the observed interlayer peak to consist of two contributions with different temperature behavior. Our calculations confirm their interpretation that the emission arises from two different kinds of interlayer excitations, one direct in momentum space [(ii)] and one indirect in momentum space [(i)]. This raises the question of the detailed nature of the intralayer scattering that assists the momentum indirect emission. Our calculations suggest a difference between the fundamental indirect and direct band gaps on the order of 140\,meV, significantly larger than the highest phonon energy in monolayer MoSe$_2$ of about 45\,meV~\cite{MoSe2-phonons}. Further insights into the relative strength and dynamics of both contributions to the interlayer exciton signal may be revealed by further time-resolved optical experiments and explicit ab initio calculations of momentum-indirect excitons with inclusion of electron-phonon interaction effects in these heterostructures.

\section{Conclusion}
To conclude, we predict that mixed MoX$_2$/WY$_2$ vertical heterostructures host two spin-orbit split Rydberg series of excitonic transitions at the $K$ point with spatial confinement of the electron and holes parts to the MoX$_2$ and WSe$_2$ layers, respectively. Their binding energy is around 250\,meV for the lowest-energy interlayer exciton. Our calculations suggest that the MoSe$_2$/WSe$_2$ heterostructure exhibits a fundamental band gap which is indirect in reciprocal space \textit{and} in real space. Taking into account the exciton binding energies, we find a second optical transition at similar energy, which is direct in reciprocal space (at the $K$ and $K'$ points of the Brillouin zone) but still indirect in real space. Such interlayer transitions (\emph{i.e.} spatially indirect) are attributed to the interlayer transitions to experimentally observed additional photoluminescence peaksand explain the asymmetric shape of the photoluminescence peaks that was recently reported to stem from a doublet of two unequal contributions.
Low oscillation strengths cause these interlayer excitons to be undetectable in absorption experiments. Our calculations show that different band alignment and orbital overlap between the constituent layers in different stacking orders have a relatively weak effect on the peak energies of the interlayer excitons but a significant effect on the oscillation strengths and their (linear) polarization dependence. The spatial separation of electrons and holes in the energetically lowest excited state makes such heterojunctions interesting for use in thin and flexible photovoltaic devices.\\

Note: During the reviewing process, two additional reports\cite{deilmann-MoS2WS2,BNMoS2WS2BN} with theoretical calculations for MoS$_2$/WS$_2$ heterostructures have been published. The results are compatible with our work. Further, a preprint on MoSe$_2$/WSe$_2$ heterostructure with results similar to ours has been uploaded to arXiv recently\cite{wirtz-arxiv}.

\begin{acknowledgments}
Computational resources used for the simulations in this work were provided by the North-German Supercomputing Alliance (HLRN) under Project bep00047. The authors acknowledge financial support by the Deutsche Forschungsgemeinschaft (DFG) within the Cluster of Excellence “Engineering of Advanced Materials”(project EXC 315) (Bridge Funding).
\end{acknowledgments}
%


\pagebreak
\clearpage
\fontsize{11}{13} \selectfont
\begin{center}
\textbf{\large Supplementary Material}\\\vspace{0.3cm}
\textbf{\large Interlayer excitons in MoSe$_2$/WSe$_2$ heterostructures from first-principles}\\
\vspace{0.6cm}
{Roland Gillen$^{1,2,*}$ and Janina Maultzsch$^{2}$}\\
\vspace{0.5cm}
{\small $^1$ Institute of Solid State Physics, TU Berlin, Hardenbergstr. 36, 10623 Berlin, Germany\\}
{\small $^2$ Department of Physics, Friedrich-Alexander University Erlangen-N\"{u}rnberg, Staudtstr. 7, 91058 Erlangen, Germany\\}
\vspace{0.2cm}
{\small $^*$ roland.gillen@fau.de\\}
\end{center}

\pagenumbering{Roman}
\setcounter{equation}{0}
\setcounter{figure}{0}
\setcounter{table}{0}
\setcounter{page}{1}
\setcounter{section}{0}
\makeatletter
\renewcommand{\theequation}{S\arabic{equation}}
\renewcommand{\thefigure}{S\arabic{figure}}
\renewcommand{\bibnumfmt}[1]{[S#1]}
\renewcommand{\citenumfont}[1]{S#1}

\section{Computational Method}\label{sec:sec1}
We used density functional theory on the level of the Perdew-Becke-Ernzerhof (PBE) approximation as implemented in the Quantum Espresso package~\citep{qe} to obtain the groundstate electronic wavefunctions and bandstructures. Integrations in reciprocal space were performed on a Monkhorst-Pack grid of 12x12x1 points in the Brillouin zone. The core electrons were replaced by normconserving pseudopotentials from the SG15 library~\citep{sg15}, where the \textit{s} and \textit{p} semi-core electrons of Mo and W are included in the set of valence electrons. Using these parameters, we optimized the atomic positions and in-plane lattice constants until the interatomic forces and the stress were below thresholds of 0.01\,eV/\AA\space and 0.01\,GPa, respectively, while keeping a vacuum layer of at least 25\,\AA\space in $c$ direction in order to minimize interactions between periodic images. 
We included semi-empirical van-der-Waals corrections from the PBE+D3~\citep{d3-2} scheme, which yields excellent predictions of the in- and out-of-plane lattice constants of layered systems~\citep{NoteS1}.

We then employed the YAMBO code~\citep{yambo} to compute the dielectric functions including electron-hole interactions and spin-orbit interaction by solving the Bethe-Salpeter Equation (BSE). We used a discrete grid of 21x21x1 k-points, 20 valence bands and 20 conduction bands for calculations including spin-orbit interaction, while denser 33x33 k-point grids, 10 valence bands and 10 conductions bands were used for calculations without spin-orbit coupling. {All BSE calculations included local field effects up to an energy of 200\,eV and a cutoff of 600\,eV for the exchange interaction, which we found sufficiently converged for our purposes.

In order to obtain an accurate description of the electronic structure, we corrected the DFT bands by quasi-particle energies from G$_0$W$_0$ calculations. We here followed the approach from Ref.~\citep{GW-extrapolation} by exploiting the linear dependence of quasiparticle corrections with the cutoff energy, which is nearly identical for difference k-point samplings.
We first calculated the quasi-particle energies using a cutoff energy of 200\,eV. We then obtained corrections for each band and k-point by performing a series of calculations using a coarser grid (of 9x9 k-points), varying the cutoff energies for the dielectric screening and the exchange contribution from 200\,eV up to 320\,eV, and extrapolating the linear dependence of the quasiparticle corrections to the limit of infinitely large cutoff energy. Electronic bandstructures were then calculated through Wannier interpolation.

The static dielectric screening and the correlation part of the GW corrections were calculated using 1500 unoccupied bands (600 bands were used for the static screening for the BSE calculations). For both GW and BSE calculations, the contribution of transitions to higher bands was included through the effective energy technique as proposed by Bruneval and Gonze~\citep{bg-eet} and represented by a single pole at an energy of 40.82\,eV. We found it important to treat the quasi-2D Coulomb interaction in our studied systems for both BSE and GW calculations using the method described in Sec.~\ref{sec:coul_treatment} in order to obtain accurate results.

\section{Modifications to the YAMBO Code}
In the following we lay out the modifications to the current version of YAMBO (4.1) we deemed necessary for our calculations.
\subsection{Proper treatment of quasi-2D Coulomb interaction and dielectric function}\label{sec:coul_treatment}
A practical difficulty when computing GW quasiparticle energies or optical spectra of two-dimensional materials arises from the long-range Coulomb interactions: due to periodic boundary conditions used in most plane-wave codes, a large vacuum layer is needed in order to obtain converged results. Another option is to use a modified Coulomb potential that truncates the interaction in the non-periodic directions. YAMBO employs a box cutoff scheme~\citep{box-cutoff} with a Fourier-transformed Coulomb potential that for two-dimensional systems takes the form
\begin{widetext}
\begin{equation}
V_c^{\mbox{2D box}}(\mathbf{q},\mathbf{G})\propto \sum_{\mathbf{q}^{'},\mathbf{G}^{'}}\frac{1}{|\mathbf{q}^{'}+\mathbf{G}^{'}|^2}\delta(\mathbf{G}_x-\mathbf{G}^{'}_x)\delta(\mathbf{G}_y-\mathbf{G}^{'}_y)\delta(\mathbf{q}_x-\mathbf{q}^{'}_x)\delta(\mathbf{q}_y-\mathbf{q}^{'}_y)F(\mathbf{q}_z+\mathbf{G}_z,\mathbf{q}^{'}_z+\mathbf{G}^{'}_z)\label{eq:eq1}
\end{equation}
with
\begin{equation}
F(\mathbf{q}_z+\mathbf{G}_z,\mathbf{q}^{'}_z+\mathbf{G}^{'}_z)=\prod_i \frac{2\sin(\mathbf{q}_z+\mathbf{G}_z-\mathbf{q}^{'}_z-\mathbf{G}^{'}_z)}{\mathbf{q}_z+\mathbf{G}_z-\mathbf{q}^{'}_z-\mathbf{G}^{'}_z}
\end{equation}
\end{widetext}

The singularity of the head of the Coulomb potential $V_c(\mathbf{q},\mathbf{G})$ for $\mathbf{q}\rightarrow 0, \mathbf{G}\rightarrow 0$ is treated my by averaging the $\frac{1}{|\mathbf{q}^{'}+\mathbf{G}^{'}|^2}$ term in the small Brillouin zone around the $\Gamma$-point through the \textit{random integration method} (RIM) as implemented in YAMBO. As within the random phase approximation, the dielectric function is given by 
\begin{equation}
\epsilon_{\mathbf{G}\mathbf{G}^{'}}(\mathbf{q})=\delta_{\mathbf{G}\mathbf{G}^{'}}-V_c(\mathbf{q},\mathbf{G})\chi_{\mathbf{G}\mathbf{G}^{'}}(\mathbf{q},\mathbf{G}),
\end{equation}
treating the divergence of the Coulomb potential around $\Gamma$ successfully restores the well-known limit
\begin{equation}
\lim_{\mathbf{q}\rightarrow \mathbf{0}}\epsilon^{-1}_{\mathbf{G}\mathbf{G}^{'}}(\mathbf{q})=1
\end{equation}

Truncation of the Coulomb potential in the non-periodic direction causes a rigid red-shift of the whole absorption spectrum compared to the bare, non-truncated case. To illustrate this, Fig.~\ref{fig:coulomb_truncation}~(a) shows calculated absorption spectra of a monolayer of boron nitride~\citep{NoteS2}
with and without truncation of the Coulomb interaction in the perpendicular direction to the layer.
\begin{figure}[b]
\centering
\includegraphics*[width=0.99\columnwidth]{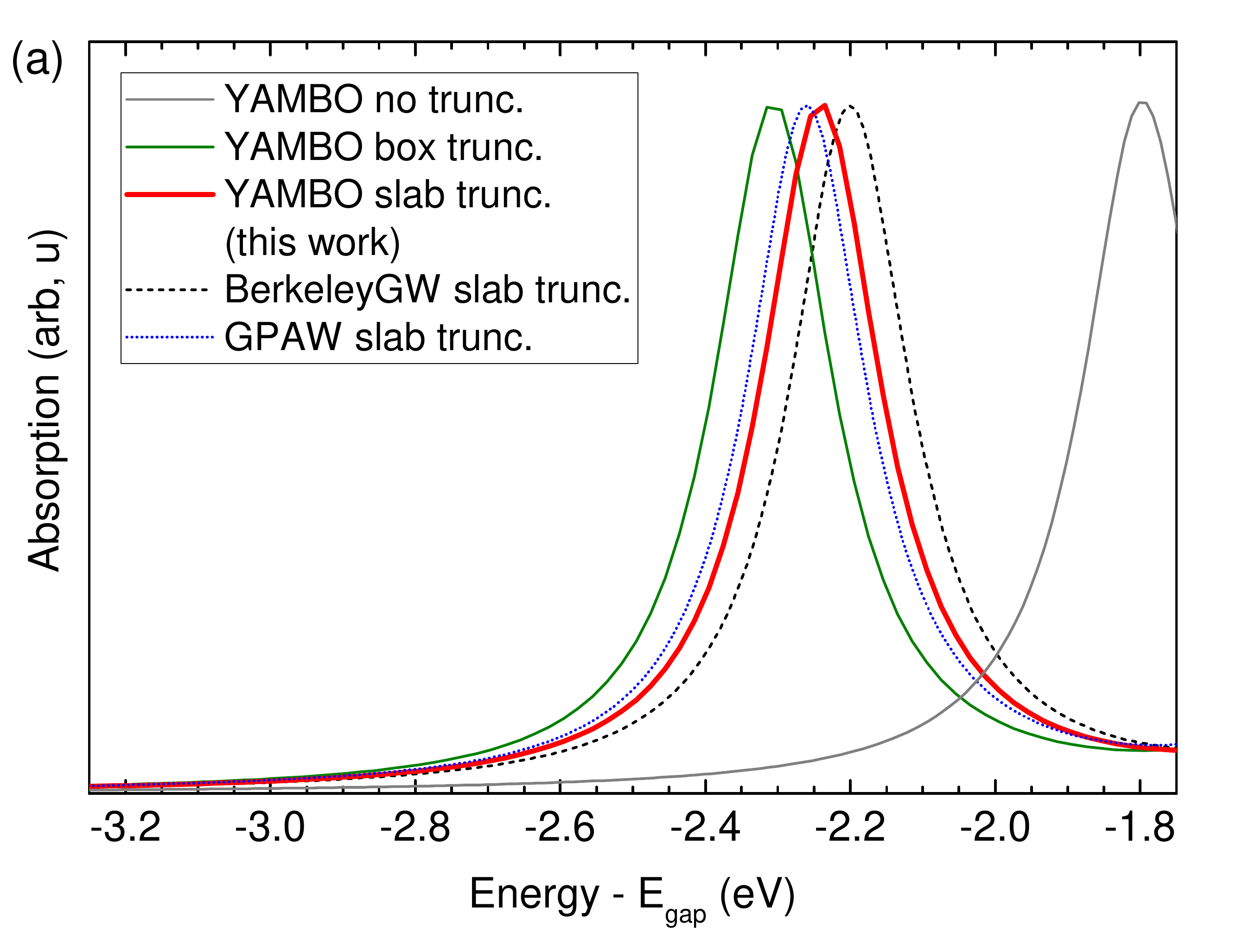}
\includegraphics*[width=0.99\columnwidth]{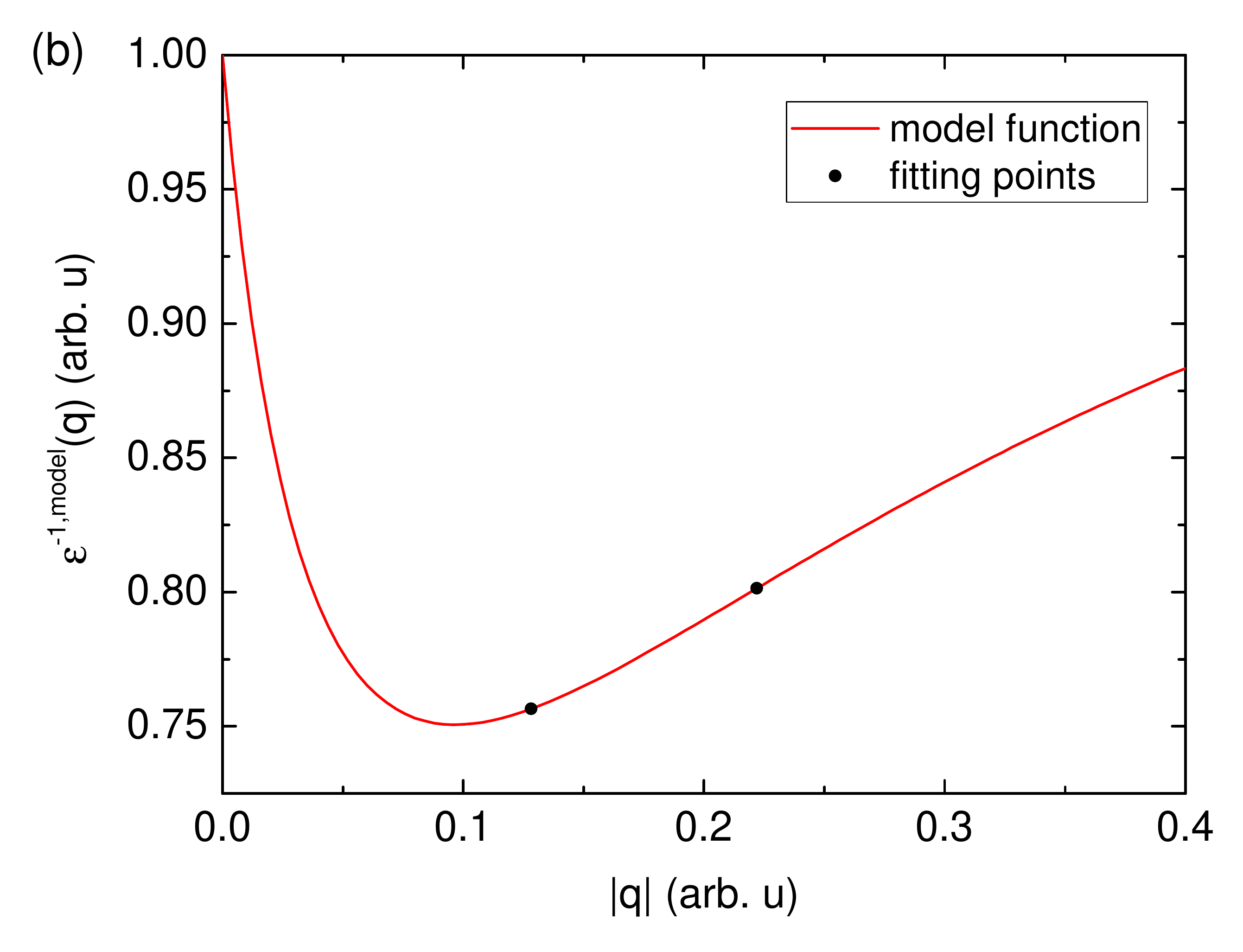}
\caption{\label{fig:coulomb_truncation} (Color online) (a) Absorption spectra of monolayer boron nitride with and without truncation of the Coulomb interaction for various codes. For a better comparability between codes, we shifted all spectra such that the zero-of-energy corresponds to the value of the calculated electronic band gap. Peak position and calculated binding energy depend noticeably on the code and employed truncation scheme. (b) Plot of the fitted model dielectric function for monolayer BN. The black dots indicate the two q-points used in the fitting procedure.}
\end{figure}
On the other hand, the Bethe-Salpeter Equation requires the screened Coulomb interaction 

\begin{equation}
W_{\mathbf{G}\mathbf{G}^{'}}(q,\omega=0)=\epsilon^{-1}_{\mathbf{G}\mathbf{G}^{'}}(q)\sqrt{V_c(\mathbf{q},\mathbf{G})V_c(q,\mathbf{G}^{'})}.
\end{equation} 

A problem arises in the current version of YAMBO (4.1) for two-dimensional materials: while the Coulomb potential in the small Brillouin zone around $\Gamma$ (i.e. $\mathbf{q}+\mathbf{G}\rightarrow \mathbf{0}$) is accounted for through RIM, the inverse dielectric function is approximated by $\epsilon^{-1}_{\mathbf{00}}(\mathbf{q}\rightarrow \mathbf{0})$. Hence, the well-known steep variation~\citep{huser-mos2} of $\epsilon_{\mathbf{00}}(\mathbf{q})$ close to $\Gamma$ is neglected, which leads to significant underscreening in the head of $W_{\mathbf{G}\mathbf{G}^{'}}(q,\omega=0)$. Correspondingly, the red-shift of the absorption spectrum and the exciton binding energies, if defined as the difference of electronic band gap and the excitonic peak energy, are overestimated. On the other hand, as Qiu \emph{et al.}~\citep{qiu-2016} pointed out, a correction of the head of the screened Coulomb interaction corresponds to an equal shift of the exciton continuum and the definition of the exciton binding energy as distance of exciton peak to the exciton continuum is a more robust choice. 

In order to facilitate comparability with experiment, however, we chose to correct $W_{\mathbf{G}\mathbf{G}^{'}}(q,\omega=0)$ for the spurious underscreening. In a first step, we implemented the well-known truncated Coulomb potential for 2D slabs of the form
\begin{widetext}
\begin{equation}
V_c^{\mbox{slab}}(\mathbf{q},\mathbf{G})\propto \frac{1}{|\mathbf{q}+\mathbf{G}|^2}\left[ 1 + e^{-K_{||}L_z/2}\left\{\frac{K_{z}}{K_{||}}\sin(K_{z}L_z/2) - \cos(K_{z}L_z/2) \right\}\right]
\end{equation}
\end{widetext}
where $L_z$ is the lattice constant in non-periodic direction, and $K_{||}$ and $K_{z}$ are the in-plane and the out-of-plane components of the reciprocal vectors $\mathbf{K}$=$\mathbf{q}$+$\mathbf{G}$, respectively. This expression should be equivalent for the box cutoff of Eq.~\ref{eq:eq1}. We then followed the prescription of Ref.~\onlinecite{ismail-beigi_suppl} and modeled the steep dependence of the inverse dielectric function near the $\Gamma$-point by the empirical function
\begin{equation}
\epsilon^{-1, \mbox{model}}_{\mathbf{00}}(\mathbf{q}) = \frac{V_c^{\mbox{slab}}(\mathbf{q},\mathbf{G}=\mathbf{0})}{1 + \gamma V_c^{\mbox{slab}}(\mathbf{q},\mathbf{G}=\mathbf{0})|\mathbf{q}|^2 e^{-\alpha|\mathbf{q}|}}\label{eq:eq2}
\end{equation}
The two free parameters $\alpha$ and $\lambda$ were fitted to the calculated inverse dielectric function at the two non-identical k-points closest to the $\mathbf{q}=\mathbf{0}$, which we expect to be a valid choice for sufficiently dense q-point samplings. Figure~\ref{fig:coulomb_truncation}~(b) shows the fitted Eq.~\ref{eq:eq2} for monolayer BN and a grid of 12x12 k-points. We then used this model function to obtain $W_{\mathbf{00}}(q\rightarrow\mathbf{0},\omega=0)$ by numerical integration of the product $\epsilon^{-1, \mbox{model}}_{\mathbf{00}}(\mathbf{q})V_c^{\mbox{slab}}(\mathbf{q},\mathbf{G}=\mathbf{0})$ on a two-dimensional sampling of 3000000 random q-points in the small Brillouin zone around $\mathbf{q}$+$\mathbf{G}$=$\mathbf{0}$. Here, 'small Brillioun zone' means the part of the full Brillouin zone of the system that is represented by the $\Gamma$ point due to the discrete k-point sampling in our calculations.

The correction causes a blue-shift of the spectrum that decrease with decreasing size of the small Brillouin zone and vanishes for infinite density of the k-point sampling. T}he new peak position is in good agreement with the prediction from the current version of BerkeleyGW (1.2), which utilizes the same model dielectric function to correct the head of the screened interaction but a slightly different approach: a different fitting method (a q-point $q_{\mbox{shift}}\approx \mathbf{0}$ is used) and setting $\alpha = 0$. For comparison, we also calculated the absorption spectrum of monolayer BN with the GPAW code, using the same parameters as before and slab truncated Coulomb interaction with the standard Coulomb integration scheme. While Coulomb truncation in general improves on the exciton binding energies, different approaches appear lead to significant differences between the quantitative values. 

Deviations between calculations with and without additional correction of $W_{\mathbf{00}}(\mathbf{q}\rightarrow\mathbf{0})$ should be particularly noticeable for materials with a strong variation of the inverse dielectric function over the small Brillouin zone for which the approximation with the screening at the Brillouin zone center is poor. For example, the averaging procedure significantly decreases the predicted binding energy of the \textsl{A} exciton of monolayer WSe$_2$ from 0.61\,meV (box cutoff w/o treatment of $W_{\mathbf{00}}(\mathbf{q}\rightarrow\mathbf{0})$) to 0.46\,meV (slab cutoff with treatment of $W_{\mathbf{00}}(\mathbf{q}\rightarrow\mathbf{0})$). Similarly, for an AA'-stacked MoSe$_2$/WSe$_2$ heterostructure, the difference of the binding energy of the interlayer exciton $X_0$ is 240\,meV~\citep{NoteS3}.

We believe that these observations might explain the scattering of reported values of exciton binding energies in transition metal dichalcogenide materials, even among publications that nominally converged Brillouin zone samplings and energy cutoffs, and draw attention to the importance of a reasonable and controlled treatment of the Coulomb interaction in 2D materials for quantitatively meaningful results. A drawback of our approach is the assumption of spherical symmetry of the dielectric screening close to the Brillouin zone center, which cannot fully include effects from structural anisotropy. A superior approach would be the one proposed recently by Rasmussen \emph{et al.}~\citep{rasmussen-2016} for GW calculations, where the head, the wings and the body of the screened Coulomb (or in that case exchange) interaction is obtained from a first-order expansion of the non-interaction density response $\chi_{\mathbf{GG}}^{-1}(\mathbf{q},\omega)$ and does not involve a fitting procedure.

\subsection{Reduction of memory-demand for direct diagonalization of BSE kernel}
The calculation of excitonic wavefunctions and oscillator strengths requires an explicit diagonalization of the BSE kernel. In practice, this is problematic as the size (and memory demand) of the BSE kernel rapidly grows with increasing density of k-point sampling, the number of valence and conduction bands and inclusion of spin-orbit interaction. In the current version of the yambo code (4.1), feasibility of direct diagonalization is severely limited by the CPU-dependent form of data storage and the implementation of the diagonalization routines: Each CPU (or rather: MPI unit, if OpenMP is used) collects the whole BSE matrix (in subroutine \textit{K\_stored\_in\_a\_BIG\_matrix}) in order to cut out its part of the kernel for the SCALAPACK diagonalization. Further, each CPU collects the whole matrix of eigenvectors (with the size of the BSE kernel) after diagonalization is finished. Thus, in the currect version of the code, direct diagonalization (i) can only be performed for systems where the size of the BSE kernel is smaller than the available memory for the processing unit, and (ii) requires the same number of MPI processes as were used for the calculation of the BSE kernel. This makes direct diagonalization prohibitively expensive or even impossible (if the BSE kernel does not fit into the available memory on one HPC node).

We have thus modified the YAMBO code in the following way:\\
\begin{itemize}
\item Instead of using a CPU-dependent storage form, we collect the BSE kernel and auxiliary data (e.g. the quasiparticle energies) in a architecture-independent single file. This is achieved through collective writing using the HDF5 library. As a result, the single CPUs/MPI units does not need to know the whole BSE matrix. This removes the severe memory bottleneck in the current YAMBO version at the possible cost of increased CPU time.
\item Each CPU/MPI unit then reads its allocated part of the BSE matrix for the SCALAPACK routine from the previously written file.
\item Typically, not all nk x nv x nc (nk/nv/nc: Number of k-points/valence bands/conduction bands) eigenvalues need to be calculated, e.g. because they do not contribute to the absorption spectrum in the energy range of interest. We thus use the PCHEEVX routine instead of PZHEEV to only calculate the lowest $n_eig$ eigenvalues and eigenvectors of the spectrum, allowing for a reduction of CPU time and file size.  
\item The calculated eigenvalues $E_\lambda$, eigenvectors $A_{\lambda}^{v,c,k}$ and the residuals $\sum_{v,c,k}\left\langle\phi_{v,\mathbf{k-q}}|e^{i\mathbf{q}r}|\phi_{c,\mathbf{k}}\right\rangle A_{\lambda}^{v,c,k}|_{q\rightarrow 0}$ are collected in a second file. 
\item The YPP code was adapted to read read the necessary data from the HDF5 files, if available.
\end{itemize}

The modified files are available from the authors upon request.

\section{M{\MakeLowercase o}S{\MakeLowercase e}$_2$-WS{\MakeLowercase e}$_2$ heterostructure}

\subsection{G$_0$W$_0$ band gaps, and electronic bands from HSE06 hybrid functional}
We here show auxilliary results for the electronic bandstructure of MoSe$_2$-WSe$_2$ heterostructures. Table~\ref{tab:tab_GW} lists the obtained electronic band gaps from our GW calculations for interlayer transitions between MoSe$_2$ and WSe$_2$ and for the intralayer layer transitions within the MoSe$_2$ and WSe$_2$ sheets. Our calculations suggest that the fundamental band gap for all three stacking orders is an indirect transitions from the $K$ valley of the valence band to the $Q$ valley conduction band.
\begin{table*}
\caption{\label{tab:tab_GW} Inter- and intralayer band gaps of MoSe$_2$-WSe$_2$ heterostructures with three different stacking orders from G$_0$W$_0$ calculations. All energies are given in eV. The fundamental band gaps are indicated in bold.}
\begin{tabular}{ c | c | c | c | c }
\hline
\hline
&$\Delta E_{K\rightarrow K}^{inter}(GW)$&$\Delta E_{K\rightarrow Q}^{inter}(\mbox{GW})$&$\Delta E_{K\rightarrow K}^{intra,MoSe_2}(\mbox{GW})$&$\Delta E_{K\rightarrow K}^{intra,WSe_2}(\mbox{GW})$\\
\hline
AA  & 1.667 & \textbf{1.53} & 2.015 & 2.011\\
AA' & 1.685 & \textbf{1.512} & 1.969 & 1.994\\
AB  & 1.742 & \textbf{1.556} & 1.978 & 1.986\\
\hline
\hline
\end{tabular}
\end{table*}

\begin{figure*}
\centering
\includegraphics*[width=0.325\textwidth]{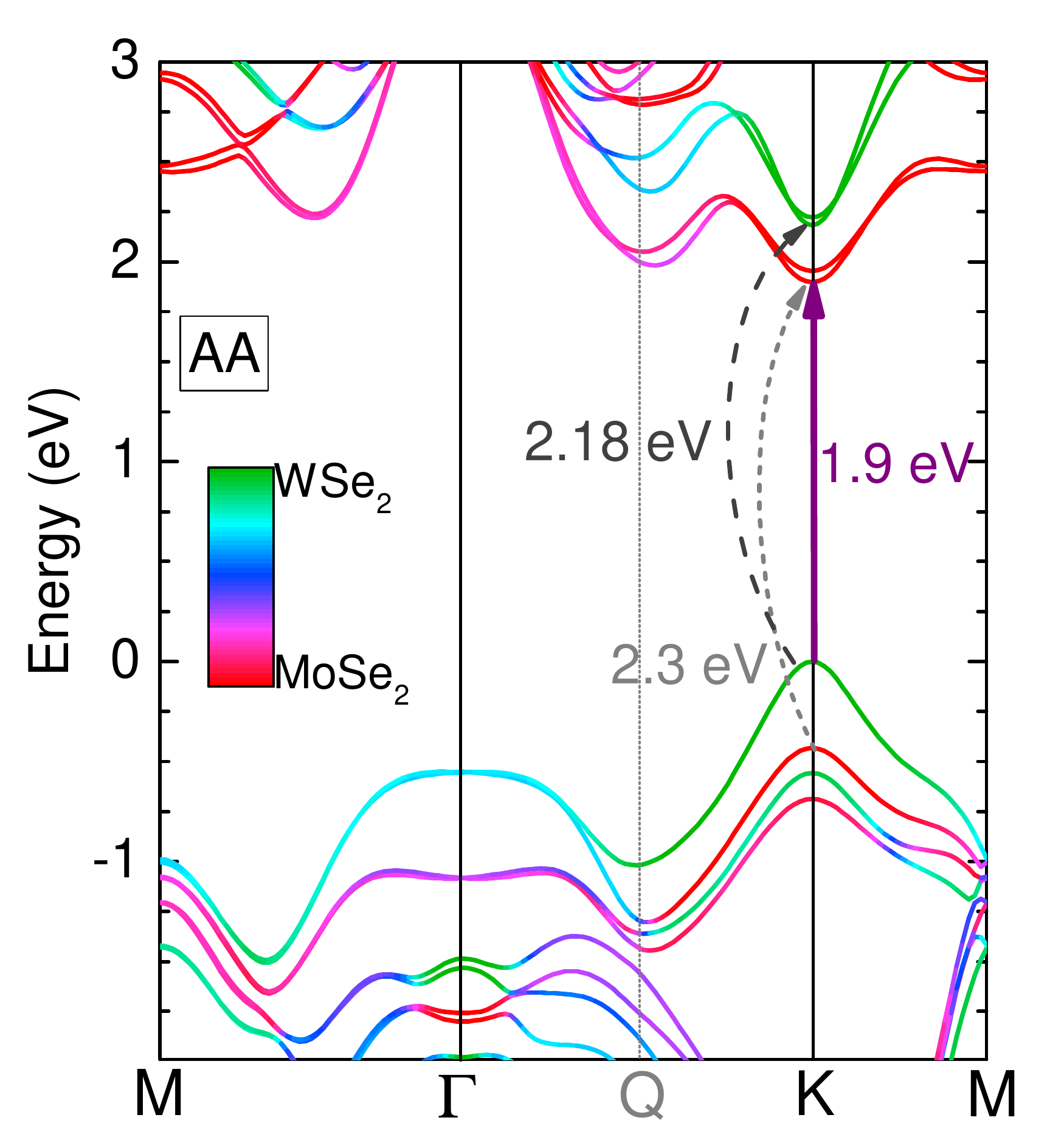}
\includegraphics*[width=0.325\textwidth]{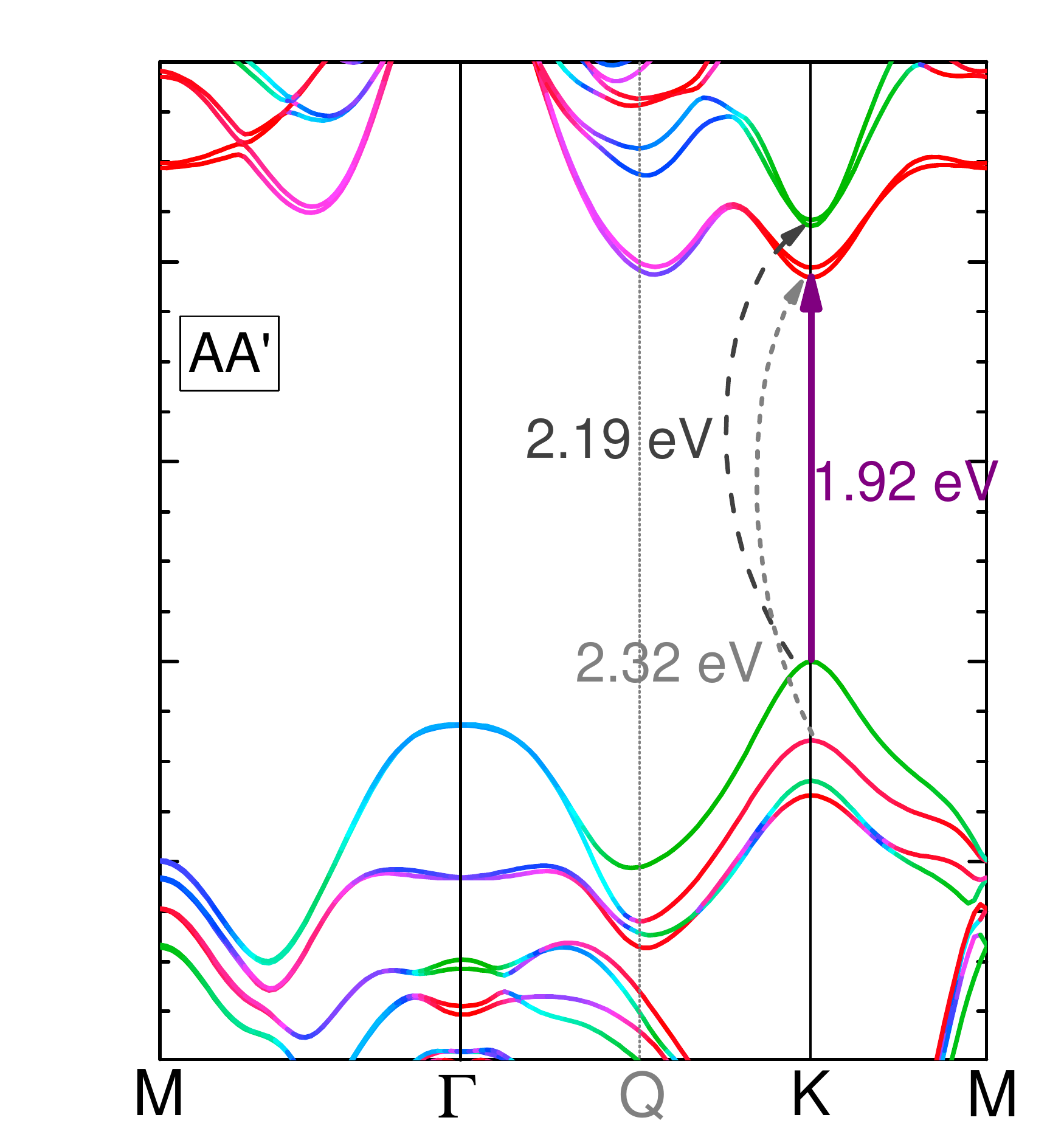}
\includegraphics*[width=0.325\textwidth]{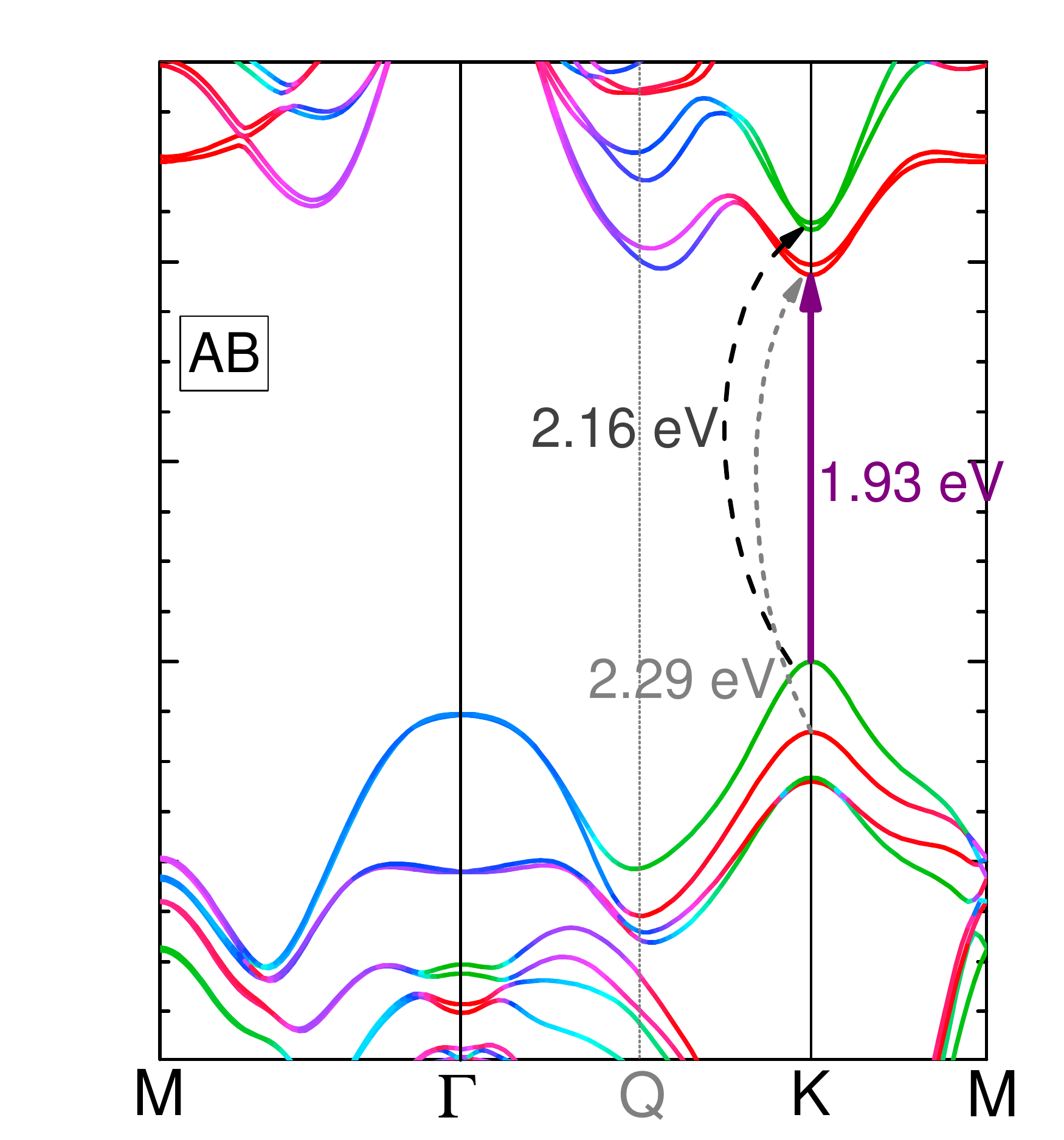}
\caption{\label{fig:MoSe2-WSe2-bands_hse06} (Color online) Electronic bandstructures for three different stackings of monolayers of MoSe$_2$ and WSe$_2$. All computations were done using the HSE06 hybrid functional and full inclusion of spin-orbit interactions. The zero-of-energy is set to the valence band maximum for each stacking. The color scale depicts the relative contributions of the materials to the bands.}
\end{figure*}

We also calculated the electronic structure from the HSE06 hybrid functional, see Fig.~\ref{fig:MoSe2-WSe2-bands_hse06} and Tab.~\ref{tab:tab_hse06}. In contrast to the G$_0$W$_0$ results, our HSE06 calculations predict the MoSe$_2$-WSe$_2$ heterostructure to possess a direct fundamental band gap for all studied stacking orders. The reason appears to be a weaker interlayer interaction near the $Q$ valley, which pushes the corresponding conduction band minimum above the conduction band valley at the $K$ point. The Coulomb divergence that appears in the exact exchange calculations have been treated by a Wigner-Seitz cutoff scheme~\citep{ws-cutoff}.

\begin{table*}
\caption{\label{tab:tab_hse06} Inter- and intralayer band gaps of MoSe$_2$-WSe$_2$ heterostructures with three different stacking orders from HSE06 hybrid functional calculations. All energies are given in eV. The fundamental band gaps are indicated in bold.}
\begin{tabular}{ c | c | c | c | c }
\hline
\hline
&$\Delta E_{K\rightarrow K}^{inter}(\mbox{HSE06})$&$\Delta E_{K\rightarrow Q}^{inter}(\mbox{HSE06})$&$\Delta E_{K\rightarrow K}^{intra,MoSe_2}(\mbox{HSE06})$&$\Delta E_{K\rightarrow K}^{intra,WSe_2}(\mbox{HSE06})$\\
\hline
AA  & \textbf{1.9}  &  1.98 & 2.3 & 2.18\\
AA' & \textbf{1.92} & 1.94 & 2.32 & 2.19\\
AB  & \textbf{1.93} & 1.97 & 2.29 & 2.16 \\
\hline
\hline
\end{tabular}
\end{table*}

\subsection{Top-view plots of excitonic wavefunctions}
\begin{figure*}
\centering
\includegraphics*[width=0.32\textwidth]{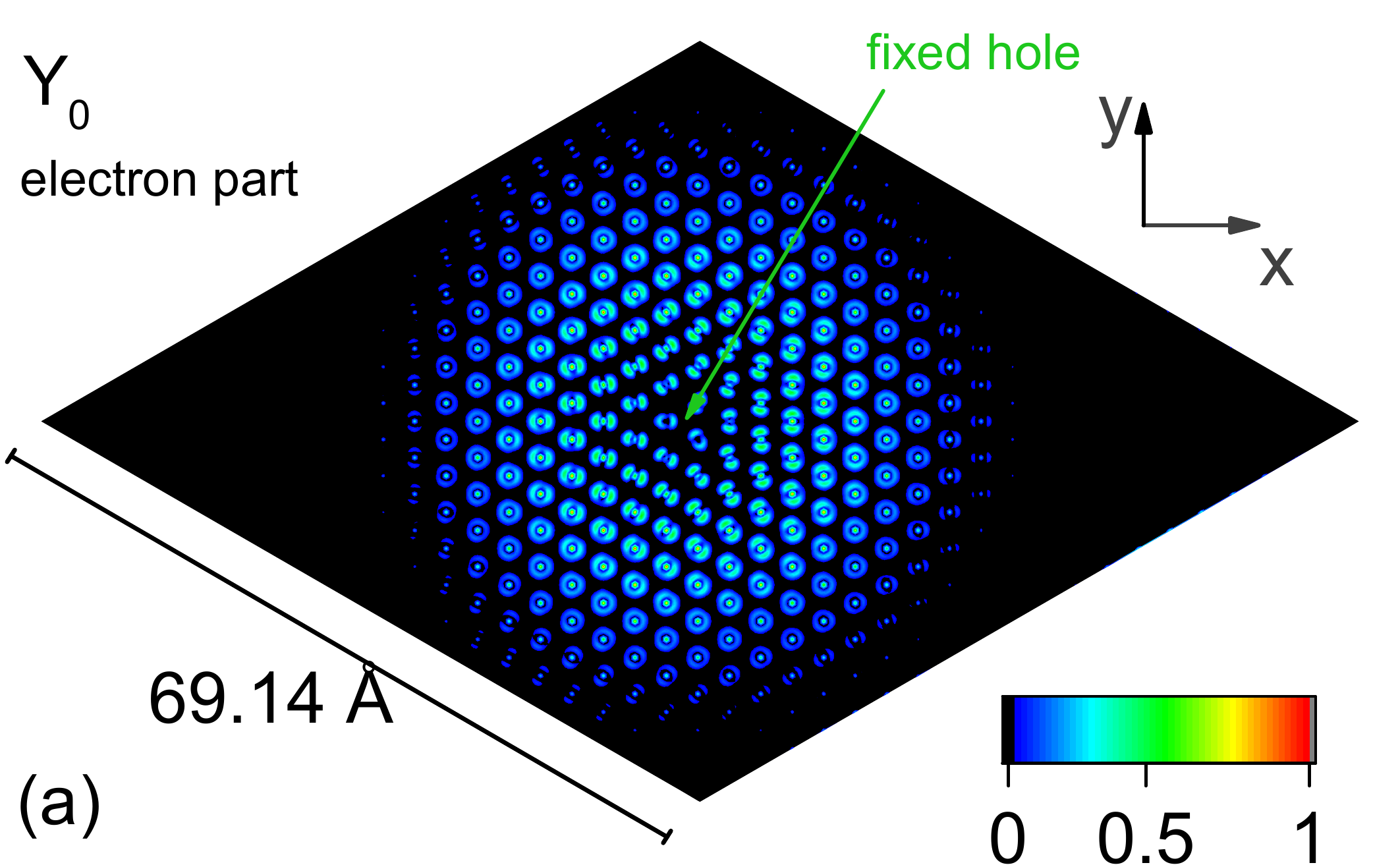}
\includegraphics*[width=0.32\textwidth]{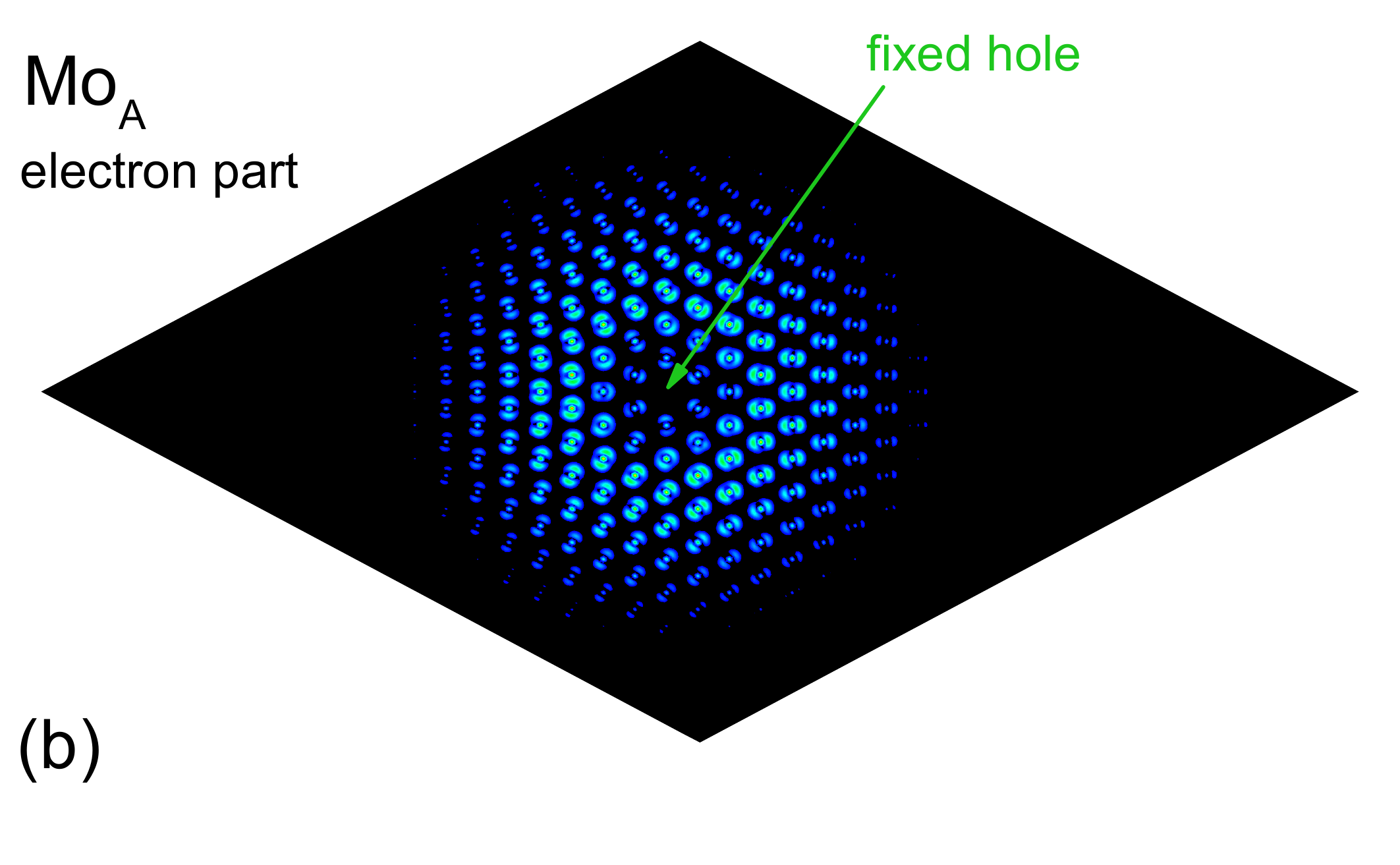}
\includegraphics*[width=0.32\textwidth]{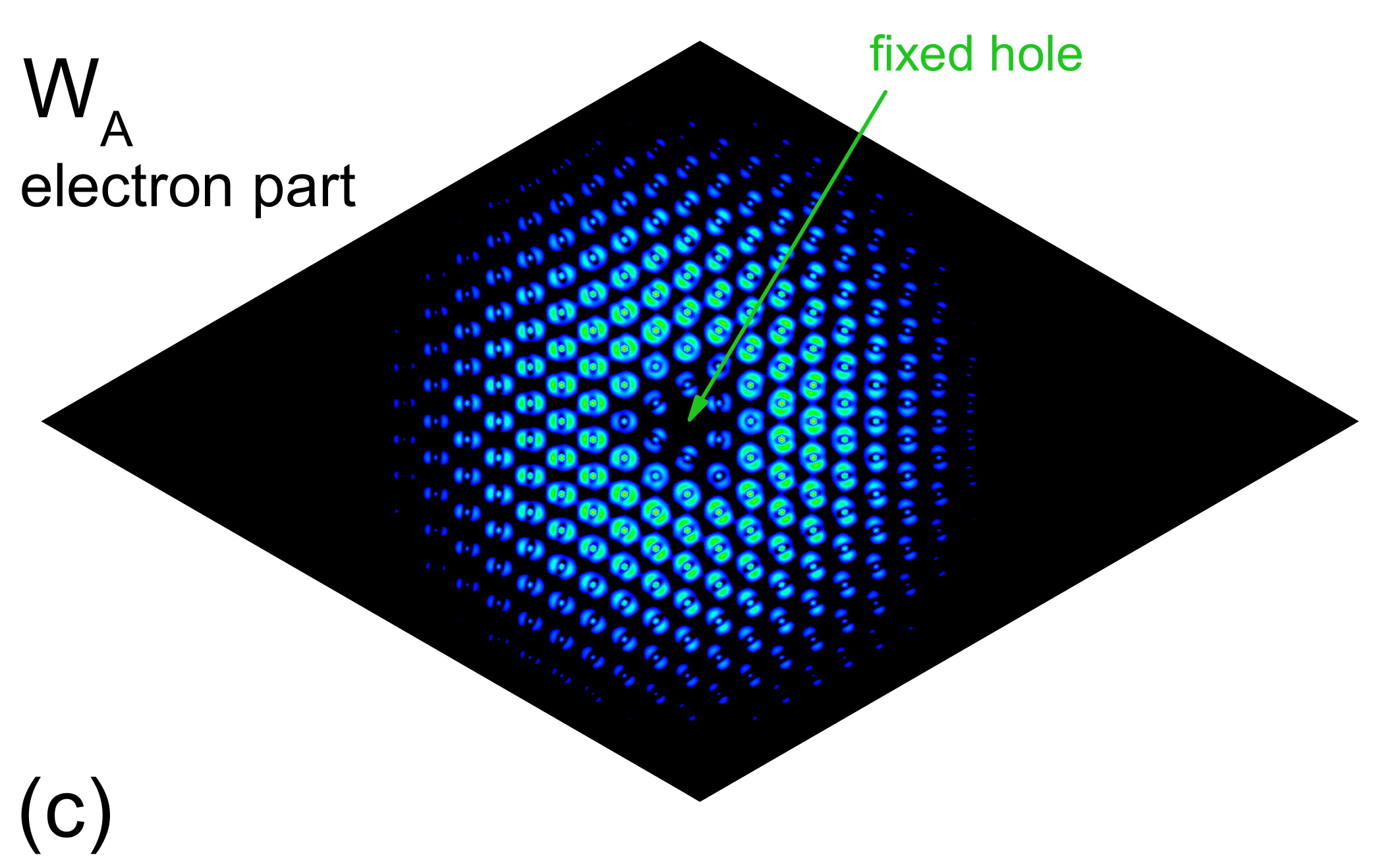}
\includegraphics*[width=0.32\textwidth]{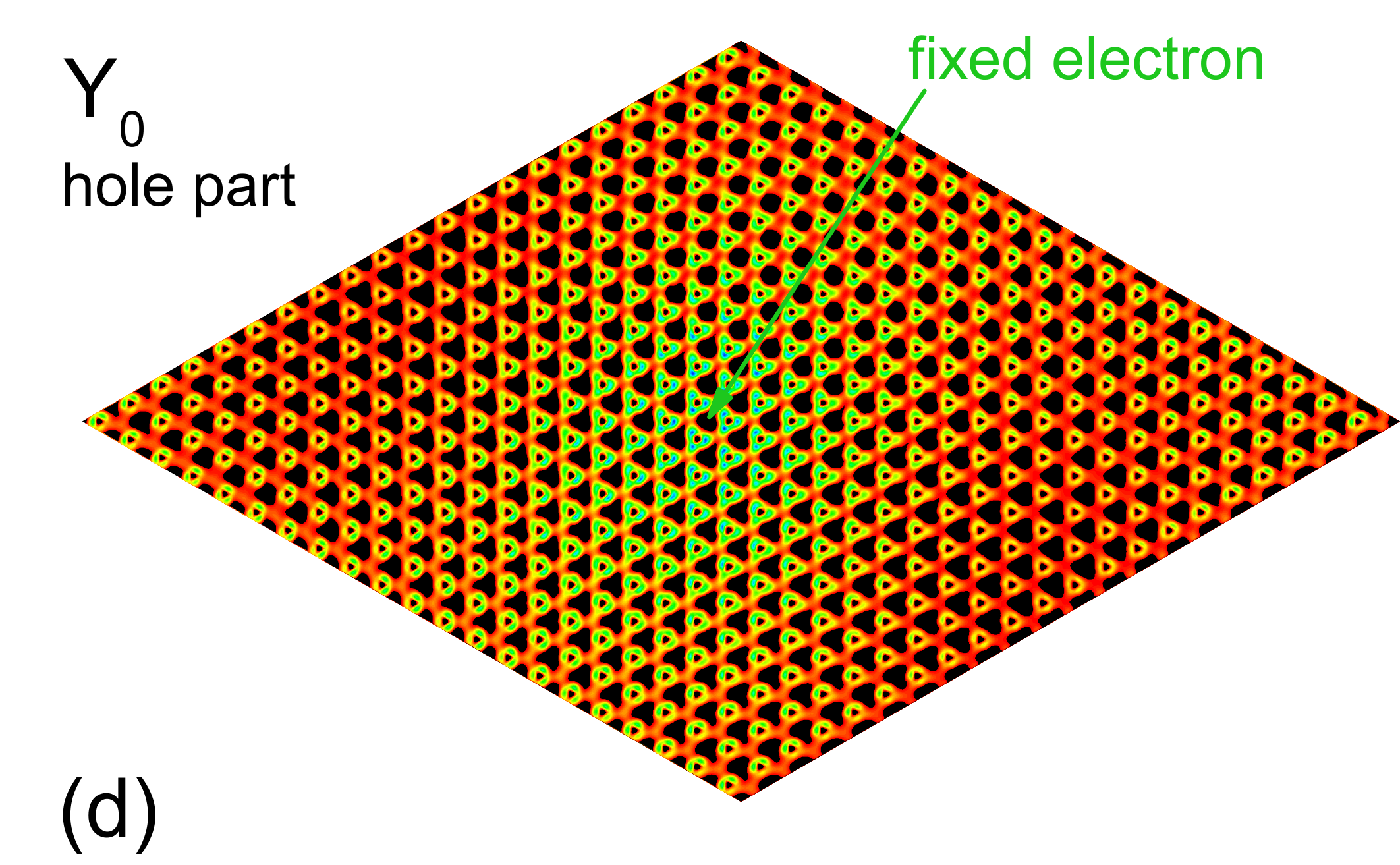}
\includegraphics*[width=0.32\textwidth]{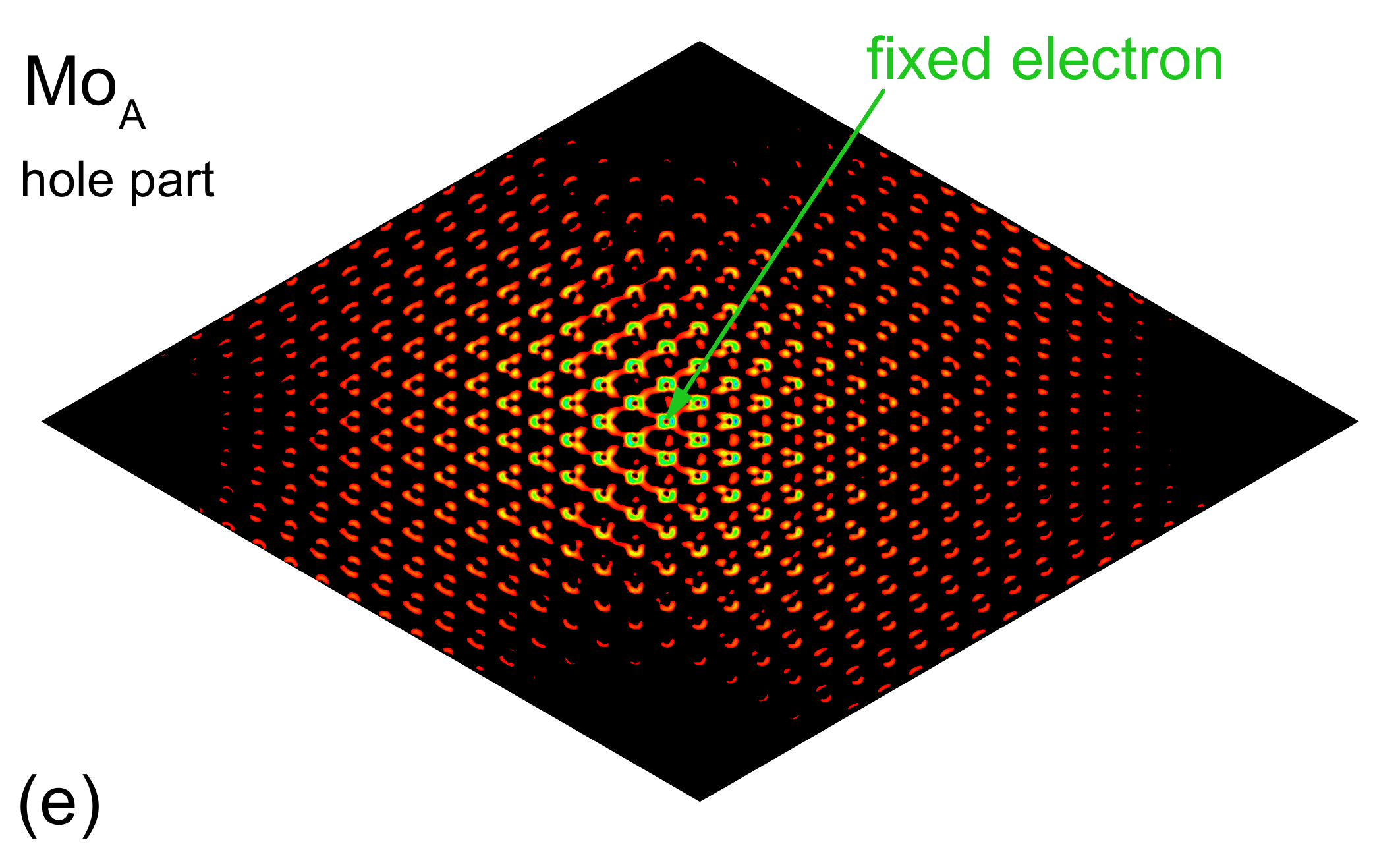}
\includegraphics*[width=0.32\textwidth]{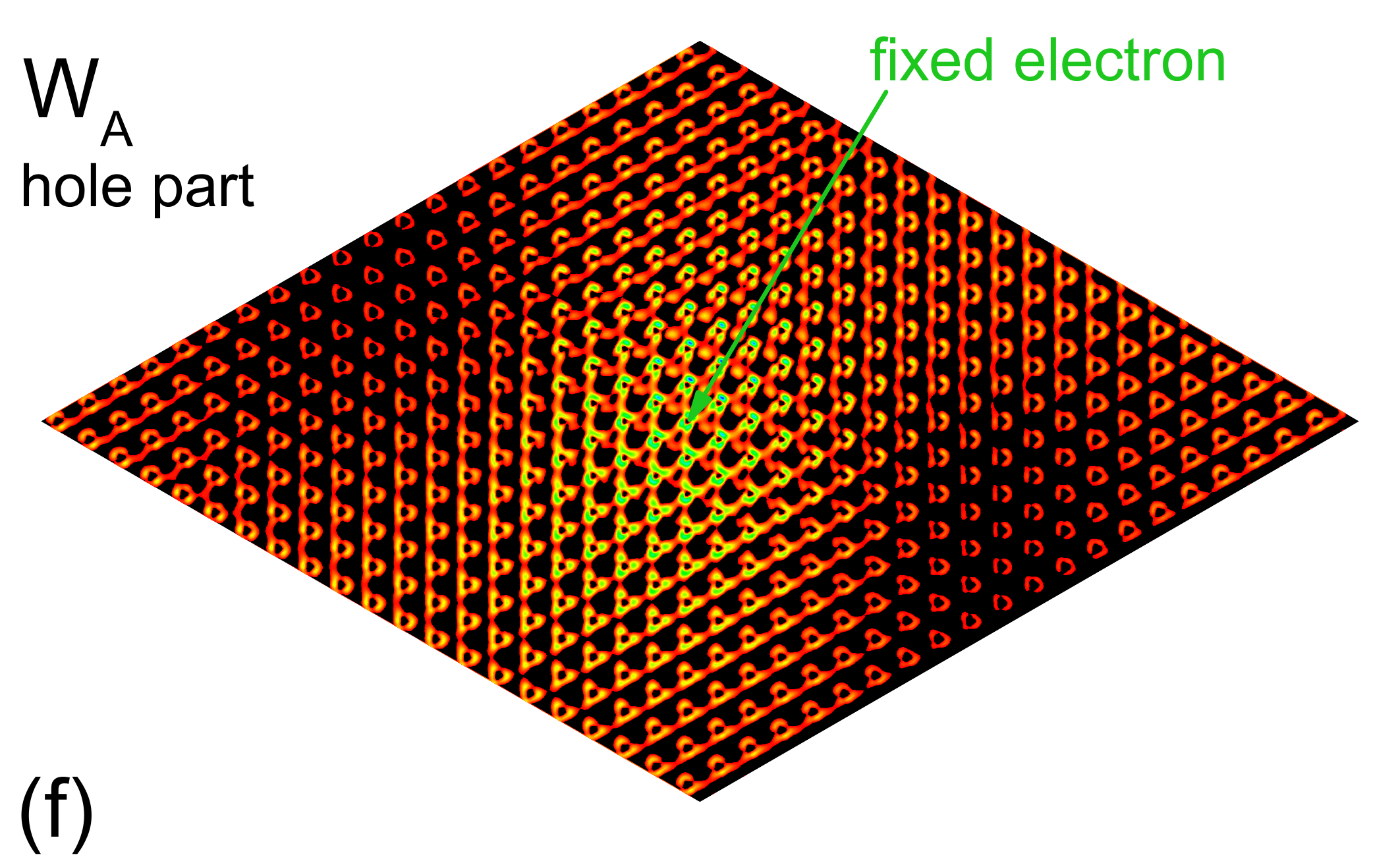}
\caption{\label{fig:exc-wfn} Top views of the excitonic wavefunctions for the \textsl{Y}$_0$, \textsl{Mo}$_A$ and \textsl{W}$_A$ transitions for AA' stacking. (a)-(c) show the electronic parts of the excitonic wavefunctions, while (d)-(f) show the hole parts. The plots are slices through the x-y planes containing the fixed hole (for electronic parts) and electrons (for hole parts), respectively. Electrons and holes were fixed on the center Mo or W atoms.}
\end{figure*}
Fig.~\ref{fig:exc-wfn} shows top views of the electron and hole parts of the excitonic wavefunctions of \textsl{Y}$_0$, \textsl{Mo}$_A$ and \textsl{W}$_A$ transitions for AA' stacking and spin-orbit coupling that correspond to Fig.~5 of the main text. For all three transitions, the electronic parts are well localized within the supercell and show a similar Bohr radius on the order of 30\,\AA. Interestingly, the electron parts exhibit a node around the location of the fixed hole that is related to the spinorial nature of the underlying electronic bands and is absent if spin-orbit coupling effects are neglected. A similar feature was found for the wavefunction of the $A$ exciton of monolayer MoS$_2$ by Molina-Sanchez~\emph{et al.}~\cite{Wirtz-mos2-excitons}.
On the other hand, the hole parts exhibit a distinct Bloch-like delocalized nature for the WSe$_2$ valence bands, which might hint at a significantly larger Bohr radius for the holes compared to the electrons and corresponding necessity of denser k-point sampling than the employed 21x21x1 grid in order to properly descibe the localization of the hole part. The difference in localization between electrons and holes and between the hole parts of \textsl{Mo}$_A$ and \textsl{W}$_A$ transitions can be traced back to the lower effective mass of the valence bands of WSe$_2$ compared to MoSe$_2$~\cite{shi-strain}.
The effect of a possible underestimation of the localization of the hole part of the excitonic wavefunction in MoSe$_2$/WSe$_2$ and similar heterostructures is subject of a future study.
\vspace{0.5cm}

\subsection{Absorption spectra for out-of-plane polarized light}\label{sec:sec1}
\begin{figure*}
\centering
\includegraphics*[width=0.32\textwidth]{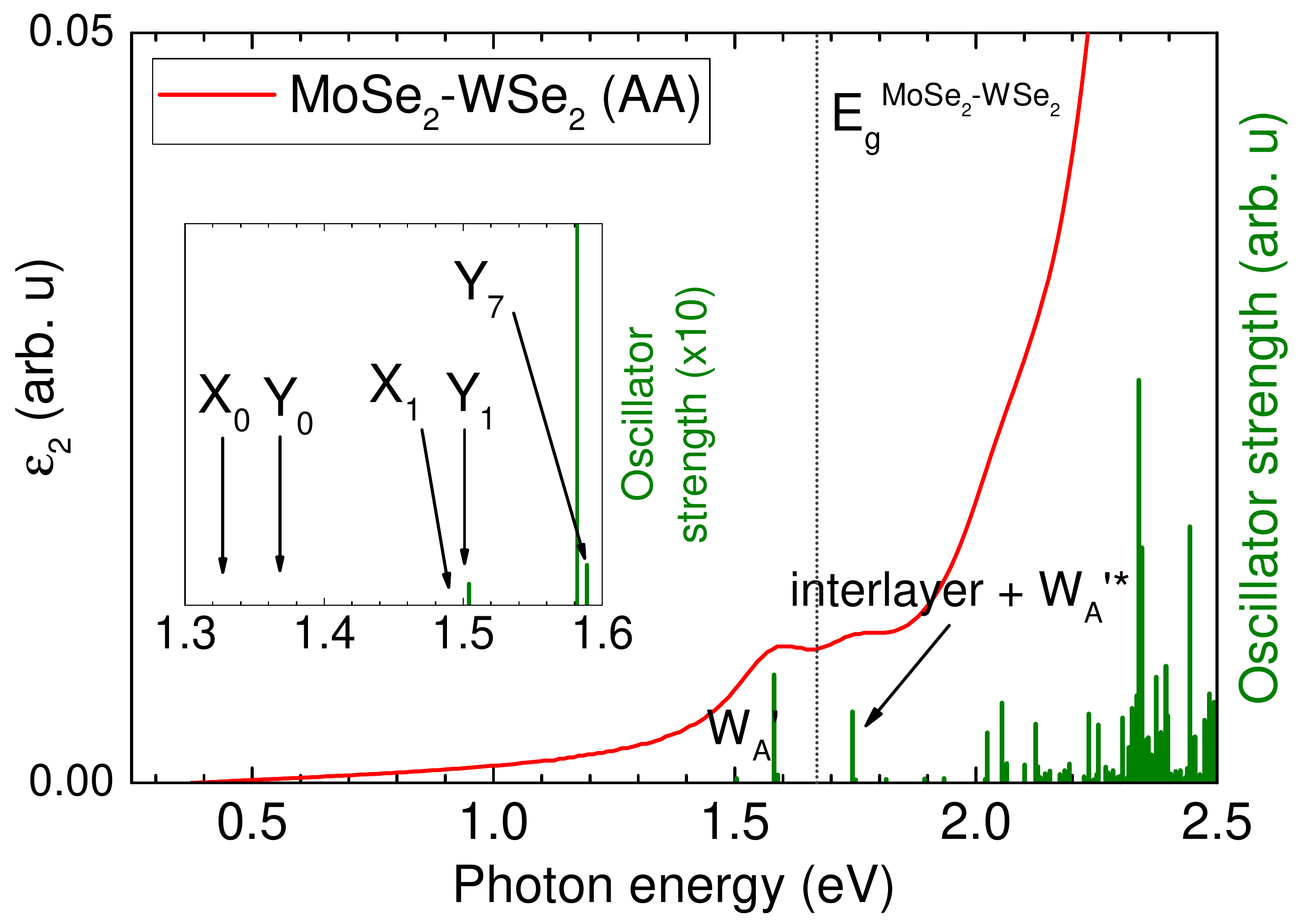}
\includegraphics*[width=0.32\textwidth]{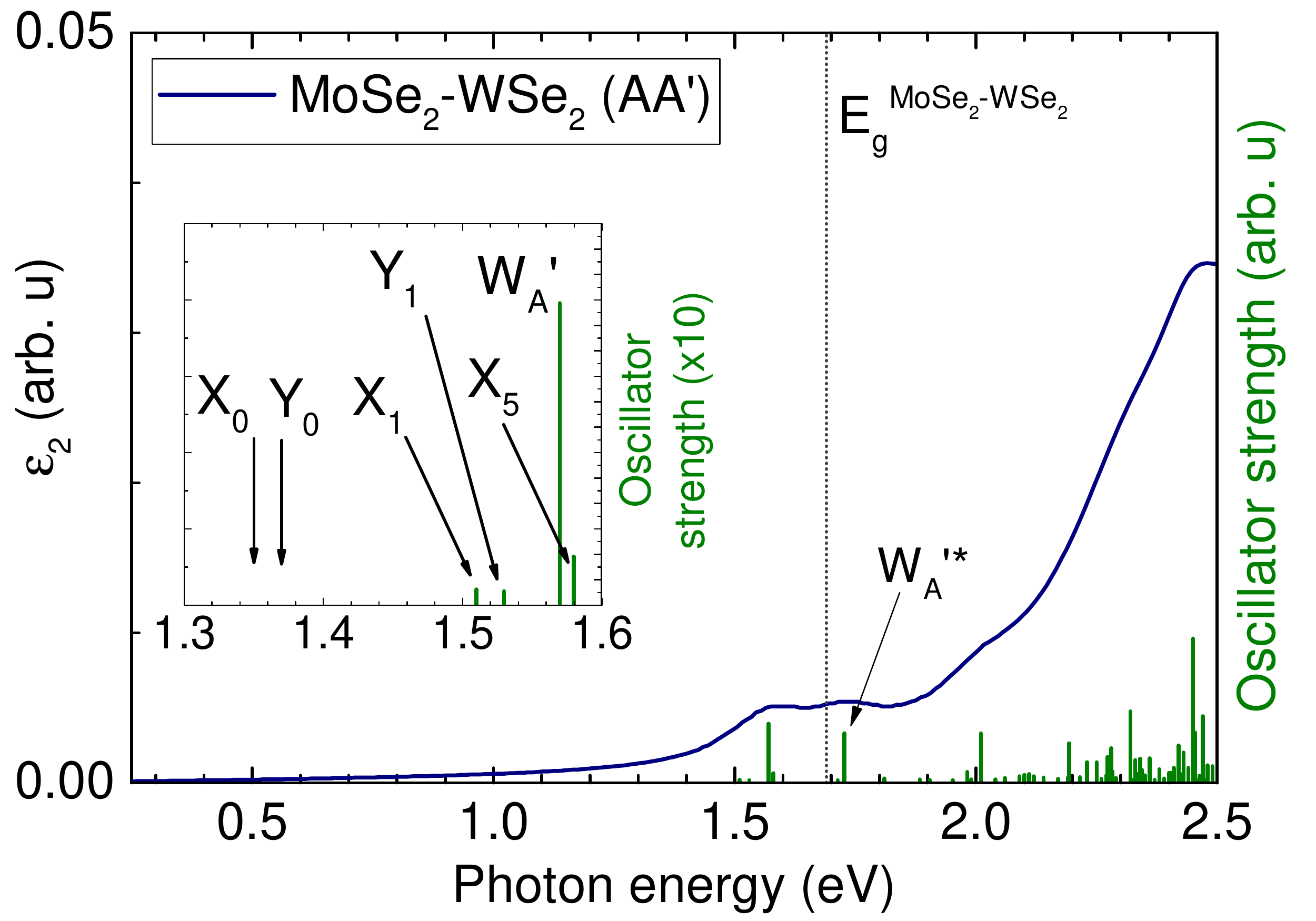}
\includegraphics*[width=0.32\textwidth]{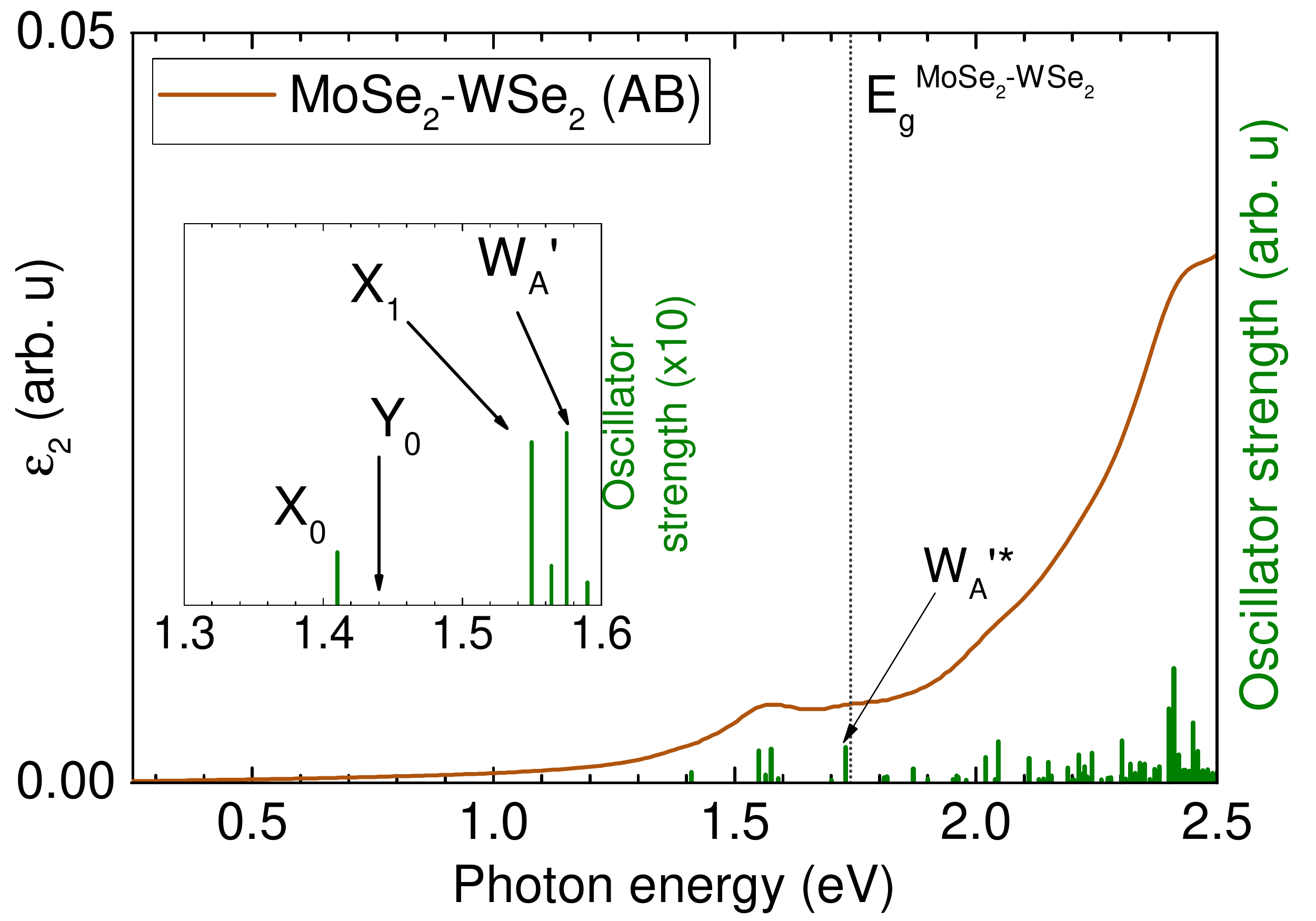}
\caption{\label{fig:absorption_perp} Same as Fig.~4 of the main test for light polarization \textit{perpendicular} to the surface of the heterostructure. Green bars show the optical oscillator strenghts of the constituting excitonic and band transitions. Note that the scale is smaller than that of Fig.~4 by a factor of 100.}
\end{figure*}
{
Fig.~\ref{fig:absorption_perp} shows the calculated dielectric functions for the three studied stacking orders for light polarization perpendicular to the 2D plane of the heterostructure. While the spectra look similar in shape as the corresponding spectra for light polarization parallel to the surface, the overall magnitude is reduced by a factor of 100 and the contribution of interlayer excitonic transitions is far more prominent. Overall, we find that, independently of the stacking order, the absorption spectrum at low energies for light polarized perpendicular to the surface is dominated by the Rydberg series of interlayer transitions and an intralayer transition in the WSe$_2$ layer (W$_A'$) and its first excitation. This intralayer transition occurs between the highest valence band and the lower of the spin-orbit split conduction bands of WSe$_2$ and is nominally forbidden by optical selection rules. We find a similar behavior for AA stacking, where the \textsl{Y}$_n$ contributions, transitions between the valence band maximum and the higher of the spin-orbit split conduction bands of the MoSe$_2$ layer are stronger than the corresponding \textsl{X}$_n$ transitions. On the other hand, the situation is reversed for AB stacking, where the \textsl{X}$_n$ contributions are significantly stronger. For AA' stacking, both \textsl{X}$_n$ and \textsl{Y}$_n$ contributions appear to be of similar magnitudes. We believe this motivates a more detailed study of optical selection rules in such stacked heterostructures of 2D TMDCs in dependence of the stacking order.

\subsection{Origin of polarization behavior of interlayer excitons}
\begin{figure*}
\centering
\includegraphics*[width=0.49\textwidth]{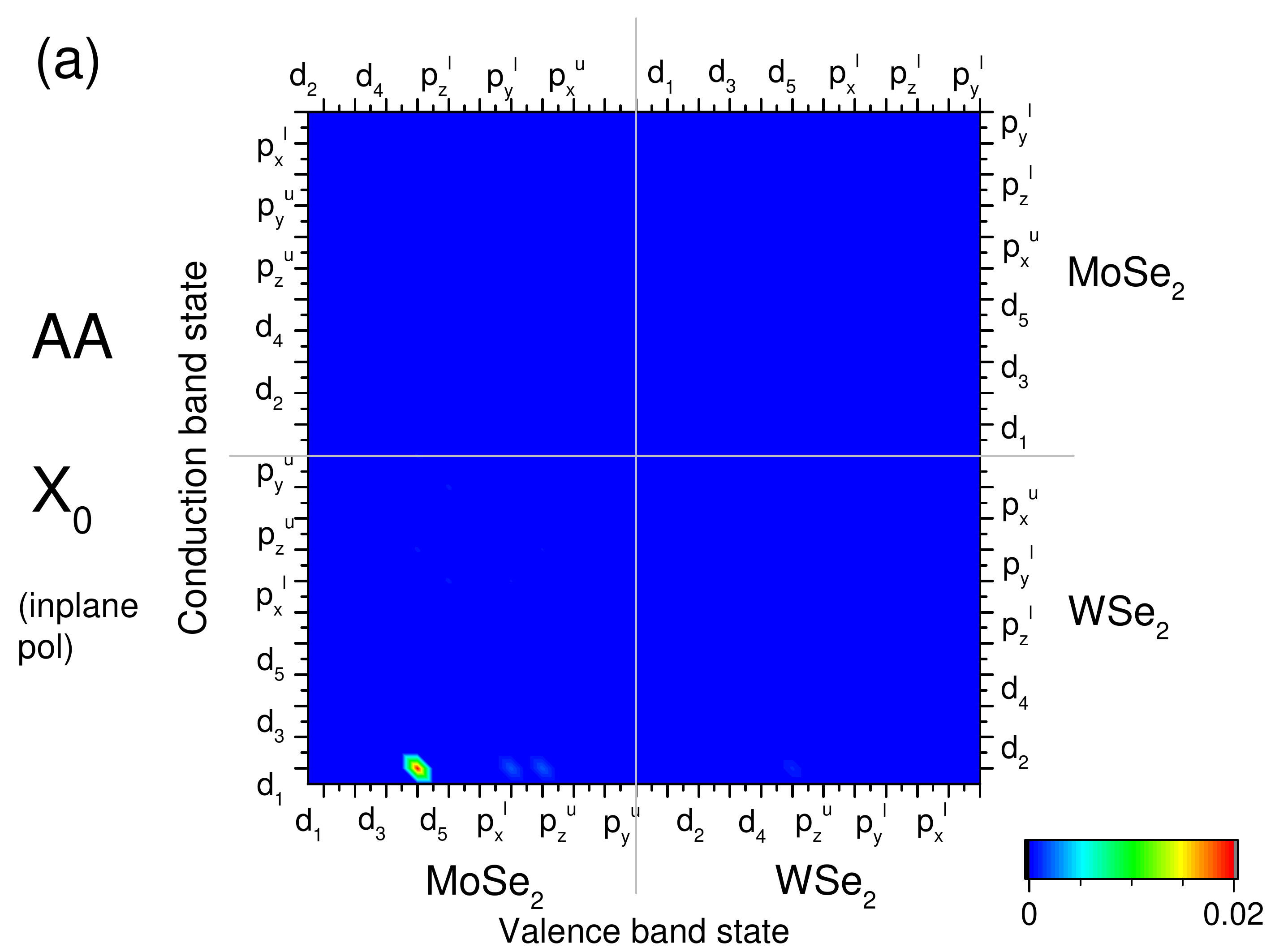}
\includegraphics*[width=0.49\textwidth]{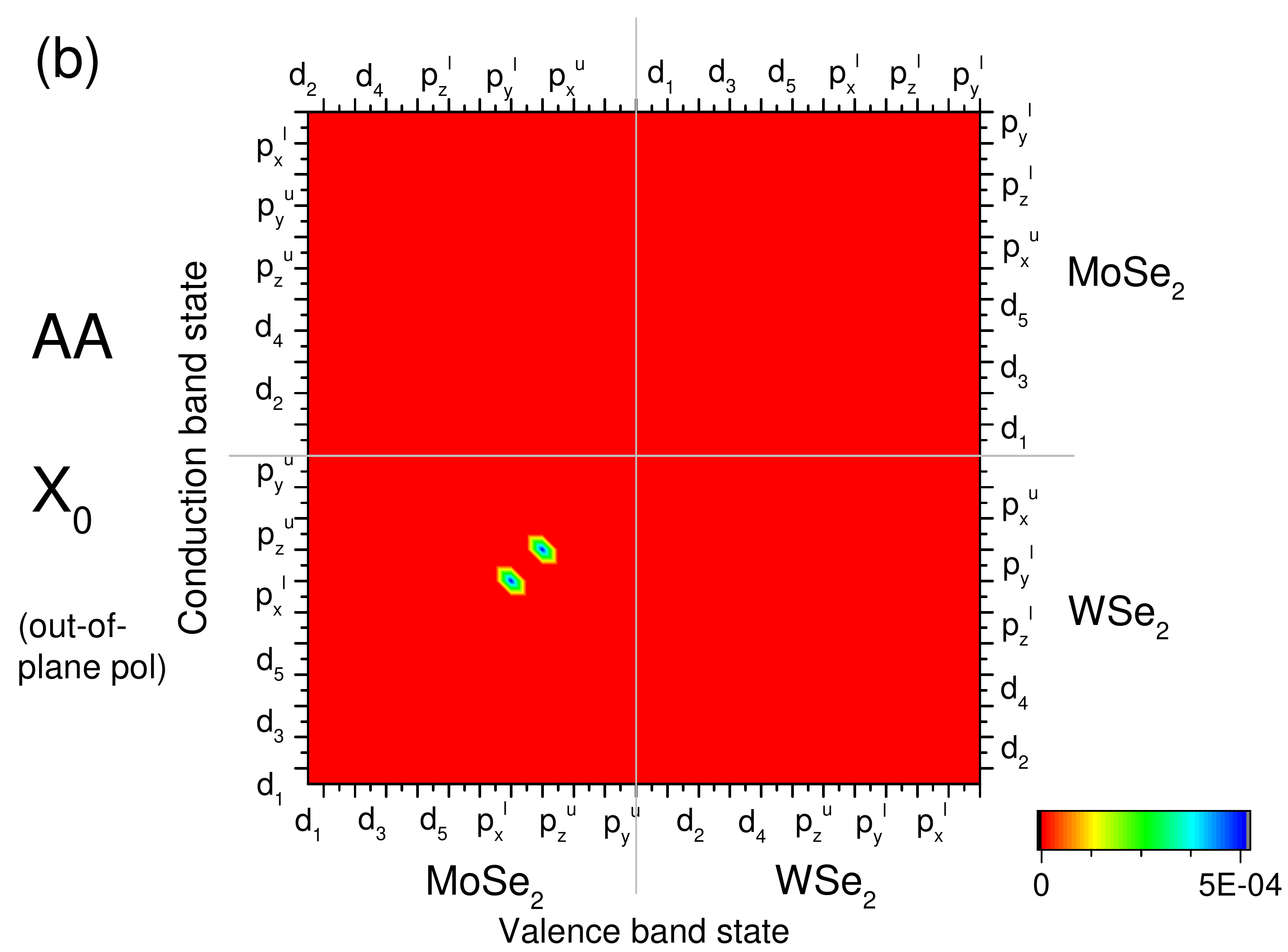}
\includegraphics*[width=0.49\textwidth]{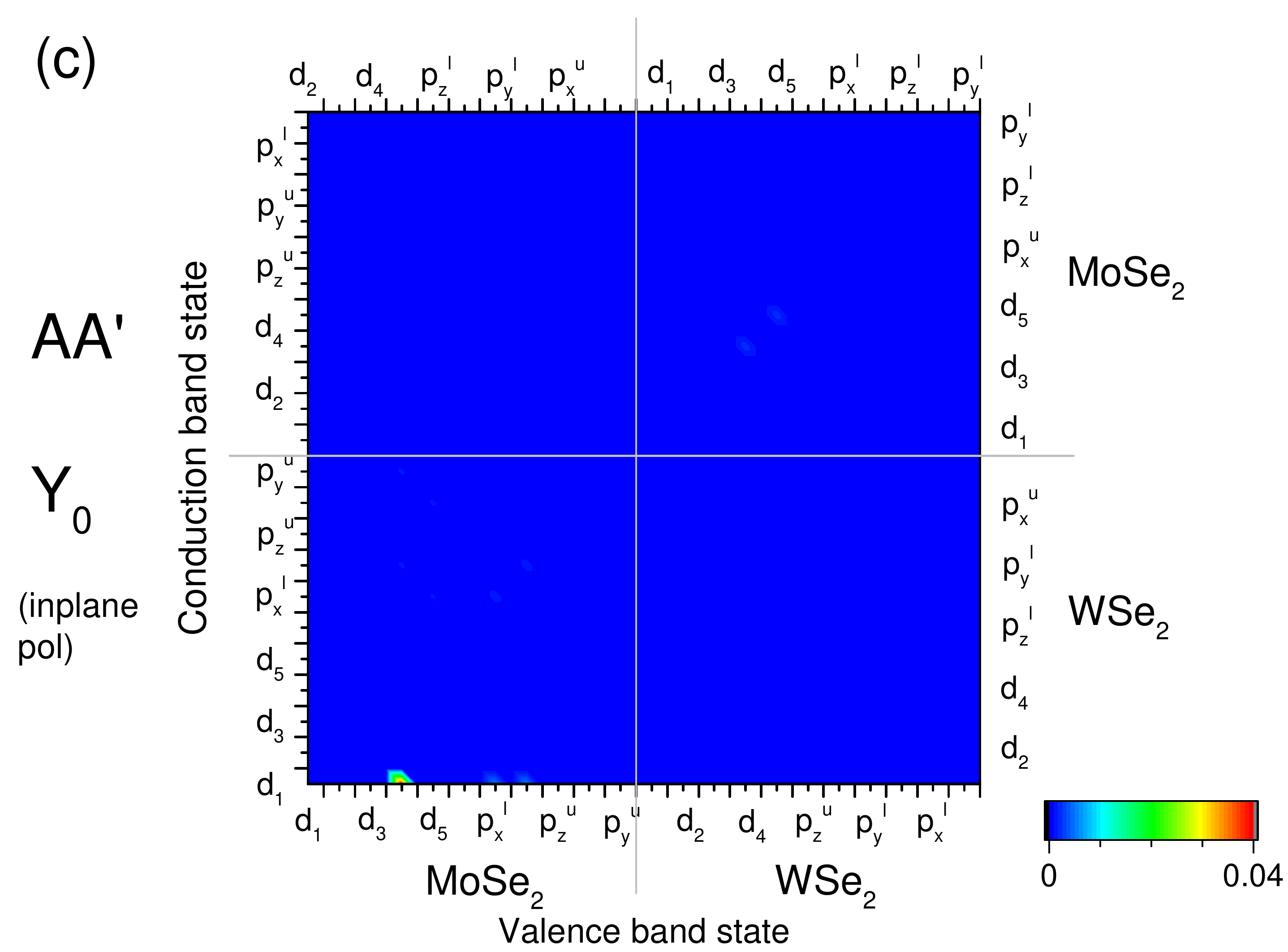}
\includegraphics*[width=0.49\textwidth]{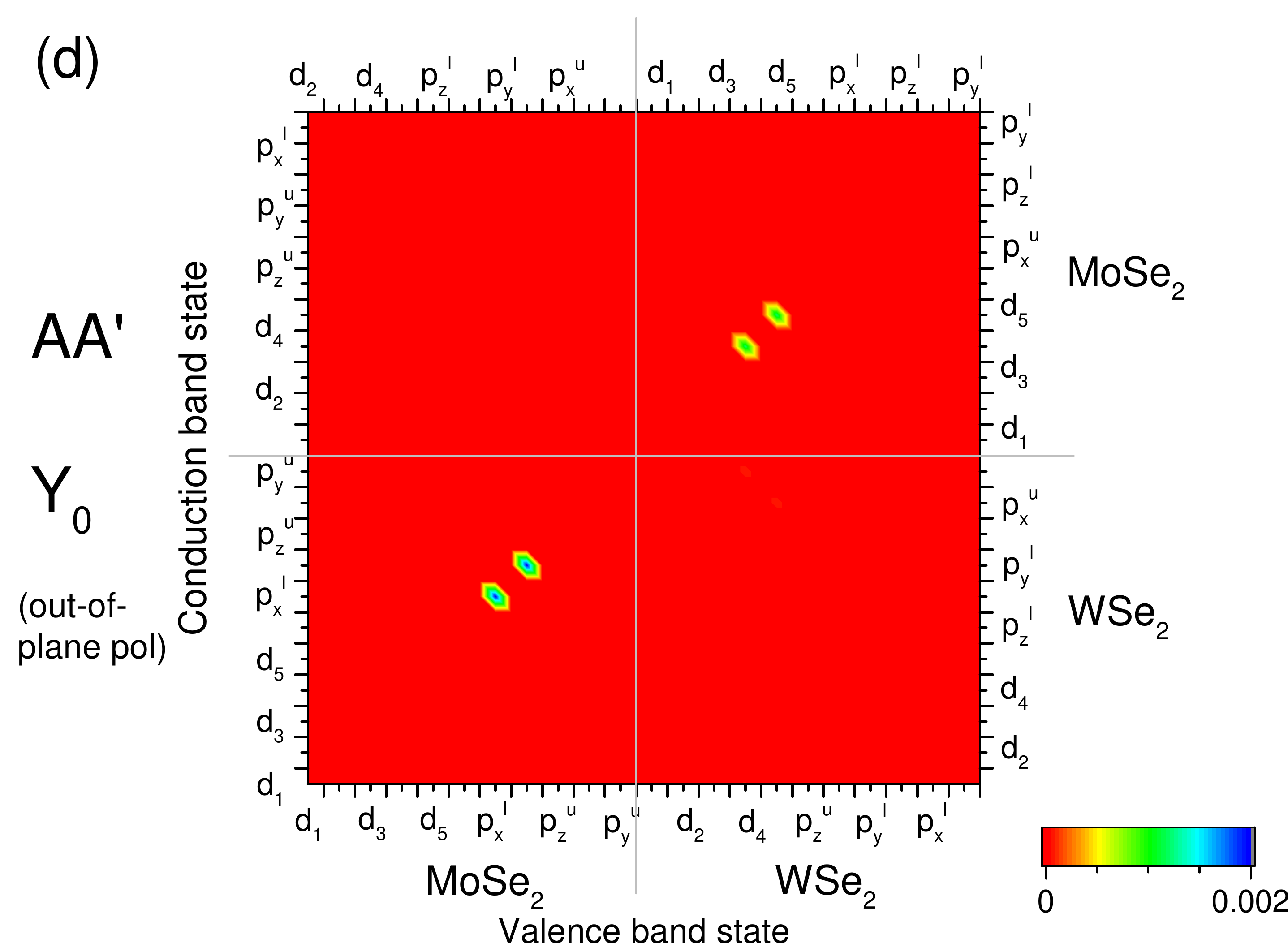}
\includegraphics*[width=0.49\textwidth]{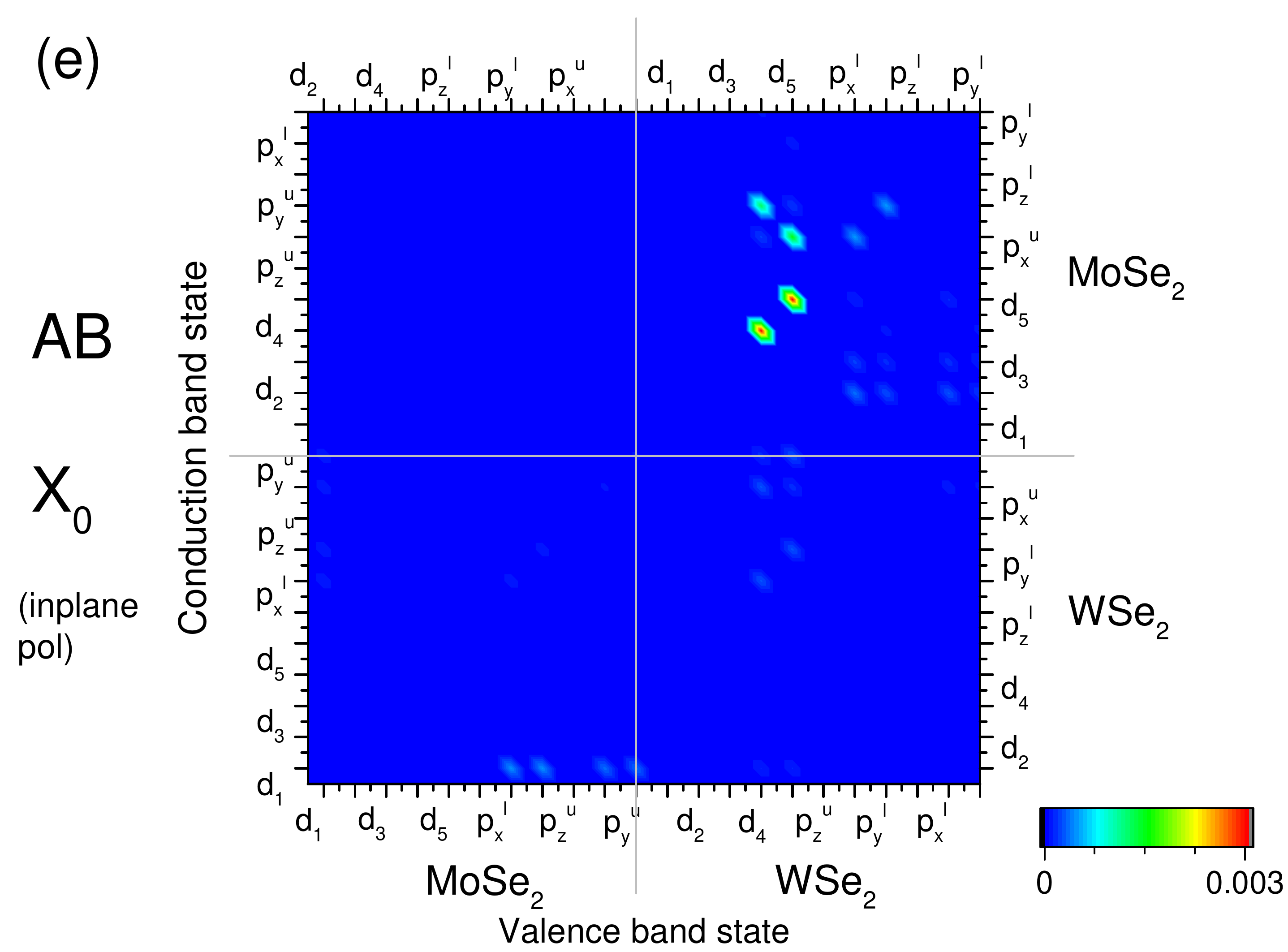}
\includegraphics*[width=0.49\textwidth]{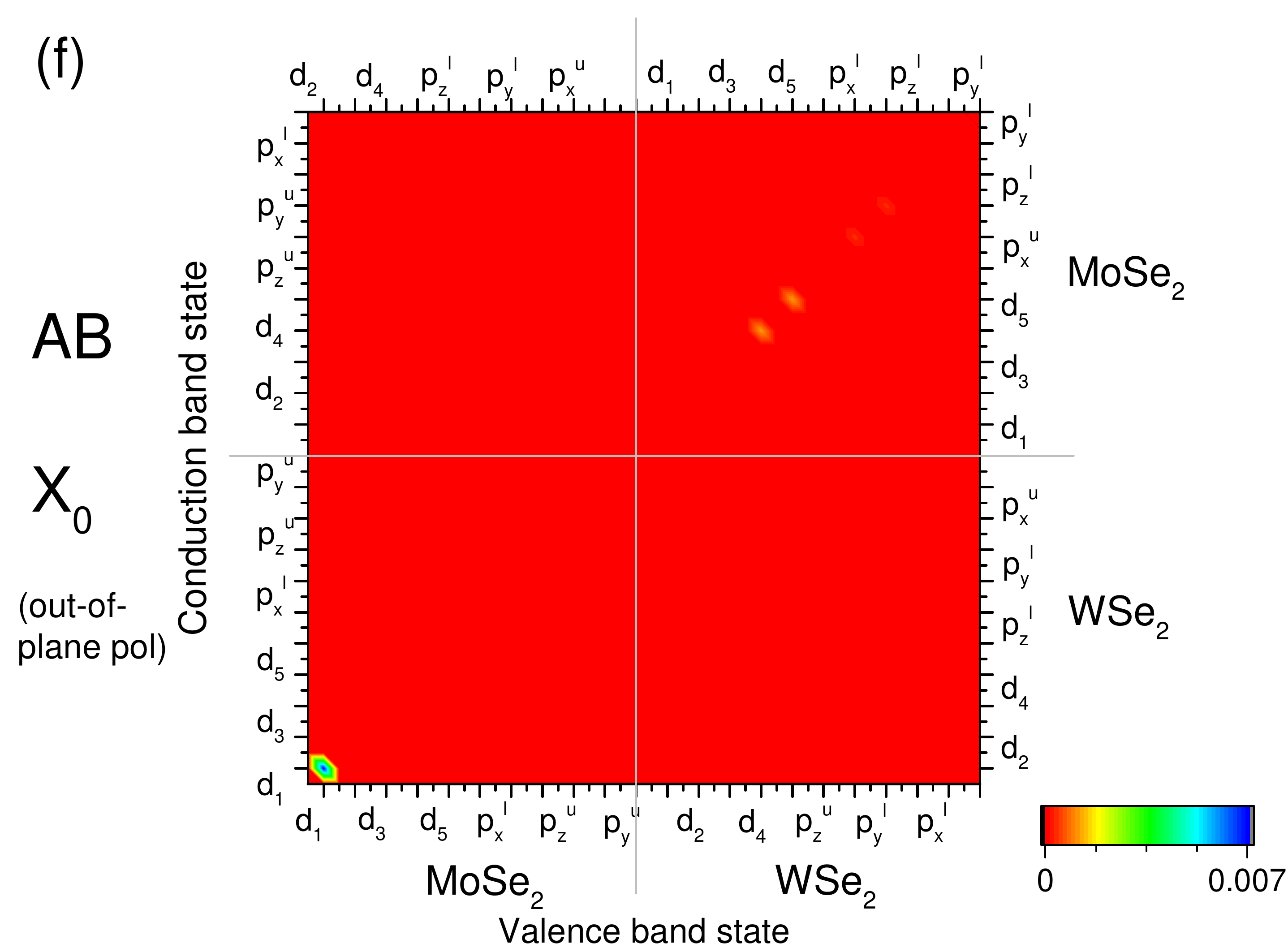}
\caption{\label{fig:exc-cont} Contributions of the transitions between atomic orbitals in the valence and conduction band to the optical matrix element of the \textsl{X}$_0$ transition (\textsl{Y}$_0$ for AA' stacking) as calculated from Eq.~\ref{eq:eq1}. (a),(c),(e) show the contributions for AA, AA' and AB stacking, respectively, for light polarization parallel to the plane of the MoSe$_2$/WSe$_2$ heterostructure, while (b),(d),(f) show the corresponding matrix elements for out-of-plane parallelization of the exciting light. For brevity, we abbreviated the $d$ orbitals. $d_1$ stands for $d_{z^2}$ orbitals, $d_2$ stands for $d_{xz}$, $d_3$ stands for $d_{yz}$, while $d_4$ and $d_5$ stand for $d_{x^2-y^2}$ and $d_{xy}$ orbitals, respectively. The indizes $l$ and $u$ stand for orbitals in the lower and upper selenium sublayers of the MoSe$_2$ and WSe$_2$ sheets. Major ticks indicate spin-down electrons, while minor ticks indicate spin-up electrons. Note that the scale can vary substantially between the plots.}
\end{figure*}

In order to obtain a better understanding of the different polarization behavior of the $X_0$ transitions for AB stacking, see also Sec.~\ref{sec:sec1} we decided to have a closer look at the optical matrix elements of the interlayer transitions. Unfortunately, untangling the differences between different stacking orders is somewhat difficult to achieve for wavefunctions in a plane-wave and reciprocal space expression of the optical matrix elements. We thus exploited the fact that Bloch states can be readily expanded in a basis of Wannier functions, \emph{i.e.}
\begin{equation}
\psi_{n\mathbf{k}}(\mathbf{r}) = \sum_{\mathbf{R}}\sum_{s} e^{i\mathbf{k}\cdot\mathbf{R}}U_{ns}w_{s\mathbf{R}}(\mathbf{r})
\end{equation}
and that the matrix elements of the expectation value of the position operator can be transformed into the same basis~\cite{yates-wannier2}:
\begin{equation}
A^W_{nm,\alpha} = \sum_{\mathbf{k}}e^{\mathbf{k}\cdot\mathbf{R}}\left\langle w_{s\mathbf{0}}|\hat{r}_\alpha|w_{t\mathbf{R}}\right\rangle,
\end{equation}
where $\alpha$ denotes the cartesian spatial directions, $\mathbf{R}$ are lattice vectors of the real space lattice and $\mathbf{k}$ are points of the employed sampling if the Brillouin zone.

We can then recover the matrix elements of the position operator in a Bloch basis by a back-transformation into the Hamiltonian gauge by the expression~\cite{yates-wannier1}
\begin{equation}
 i\left\langle \psi_{n\mathbf{k}}|\nabla_{\mathbf{k},\alpha}|\psi_{m\mathbf{k}}\right\rangle   = A^W_{nm,\alpha} + i\left(U^\dagger\partial_\alpha U\right)_{nm}\label{eq:eq1}
\end{equation}
\begin{table*}
\caption{\label{tab:contrib} Detailed list of composition of the valence and conduction band edges at the K point for three different stackings as obtained from a projection of the electronic wavefunctions onto a set of atomic orbitals. For the AA' stacking, the listed conduction band edge is the second lowest conduction band that is spin-matched with the valence band maximum due to the 180$^\circ$ relative rotation of the layers that leads to a rotation of K point of the MoSe$_2$ layer on the K' point of the WSe$_2$ layer. While the contributions of WSe$_2$ (MoSe$_2$) states to the valence (conduction) band edge does not change, the nature of the additionally mixing-in states significantly depends on the stacking order.}
\begin{tabular}{ c  c }
\hline
\hline
&Valence band edge \\
\hline
\multirow{2}{*}{AA}  & 40.6\%*(W $d_{x^2-y^2}$ + W $d_{xy}$) + 9.07\%*(Se$^{W}p_{x}$ + Se$^{W}p_{y}$)\\
    & +0.0375\%*(Mo $d_{xz}$+ Mo $d_{yz}$) + 0.21\%*(Mo $d_{x^2-y^2}$+Mo $d_{xy}$) +0.072\%*(Se$^{Mo}p_{x}$ + Se$^{Mo}p_{y}$)\\
&\\
\multirow{2}{*}{AA'}  & 40.14\%*(W $d_{x^2-y^2}$ + W $d_{xy}$) + 8.92\%*(Se$^{W}p_{x}$ + Se$^{W}p_{y}$)\\
    & +0.18\%*(Mo $d_{xz}$+ Mo $d_{yz}$) + 0.5\%*(Mo $d_{x^2-y^2}$+Mo $d_{xy}$) +0.23\%*(Se$^{Mo}p_{x}$ + Se$^{Mo}p_{y}$)\\
&\\
\multirow{2}{*}{AB}  & 40.82\%*(W $d_{x^2-y^2}$ + W $d_{xy}$) + 8.8\%*(Se$^{W}p_{x}$ + Se$^{W}p_{y}$)\\
    & +0.01\%*(Mo $d_{z^2}$) +0.019\%*(Se$^{Mo}p_{x}$ + Se$^{Mo}p_{y}$)\\
\hline
\hline
&\\
\hline
\hline
&Conduction band edge \\
\hline
\multirow{2}{*}{AA}  & 83.38\%*(Mo $d_{z^2}$) + 8.23\%*(Se$^{Mo}p_{x}$ + Se$^{Mo}p_{y}$)+0.071\%*(Se$^{Mo}p_{z}$)\\
    & \\
&\\
\multirow{2}{*}{AA'}  & 83.26\%*(Mo $d_{z^2}$) + 8.24\%*(Se$^{Mo}p_{x}$ + Se$^{Mo}p_{y}$)+0.072\%*(Se$^{Mo}p_{z}$)\\
    & \\
&\\
\multirow{2}{*}{AB}  & 83.02\%*(Mo $d_{z^2}$) + 8.11\%*(Se$^{Mo}p_{x}$ + Se$^{Mo}p_{y}$)+0.068\%*(Se$^{Mo}p_{z}$)\\
    & + 0.258\%*(W $d_{z^2}$) + 0.07\%*(W $d_{xz}$ + W $d_{yz}$) + 0.03\%*(W $d_{x^2-y^2}$ + W $d_{xy}$) \\
		& + 0.06\%*(Se$^Wp_{x}$ + Se$^Wp_{y}$)\\
\hline
\hline
\end{tabular}
\end{table*}

We calculated the optical matrix elements $\left\langle w_{s\mathbf{0}}|\hat{r}_\alpha|w_{t\mathbf{R}}\right\rangle$ and the unitary rotation matrix $U_{nm}$ and its derivative with the wannier90 code~\cite{wannier90}. While in our experience, the maximally-localized Wannier functions (MLWF) in transition-metal dichalcogenides are almost identical to atomic orbitals, we decided to avoid the influence of possible stacking-dependent differences in the calculated MLWFs by using Mo and W $d$ and Se $p$ atomic orbitals for the projection, without minimization of the spread functional (this was achieved by setting \textit{num\_iter=0} for the wannierization). We used a sampling of 15x15x1 k-points for the wannierization procedure, which corresponds to an expansion of the Bloch wavefunctions into wannier functions on a supercell of 15x15x1 unit cells in real space, and the full spinorial nature of the electronic wavefunctions. Electron-hole effects are not important for the qualitative trends of the optical matrix elements (but of course for quantitative values) and were neglected in our investigations.

Figure~\ref{fig:exc-cont} shows the contributions of transitions between atomic orbitals to the optical matrix element of the \textsl{X}$_0$ transition (\textsl{Y}$_0$ for AA' stacking) for the three studied stacking orders. Our calculations suggest that the oscillation strength of the studied interlayer transitions is dominated by contributions from transitions between Mo or W $d$ or between states. These transitions arise from a weak hybridization of valence and conduction bands of different layers that mix a tiny amount of Mo $d$ (W $d$), and to a lesser amount Se $p$, orbitals into the valence maximum (conduction band minimum) at the $K$ point. As Tab.~\ref{tab:contrib} indicates, this hybridization and the detailed 'spilling over' of orbitals depends on the stacking order and causes qualitative differences in the optical matrix elements. For AA stacking, we find that the strongest contribution comes from transitions of Mo $d_{x^2-y^2}$ mixed into the valence band maximum to Mo $d_{z^2}$ states in the valence band. On the other hand, for AB stacking, Mo $d_{z^2}$ are mixed into the valence band (and W $d_{z^2}$ states into the conduction band), while we do not have a contribution from Mo $d_{x^2-y^2}$ in the valence band. The dominant contribution to the optical matrix element between the valence and conduction band edges comes then from a transition between the small fraction of Mo $d_{z^2}$ in the valence band to Mo $d_{z^2}$ in the conduction band. This offers a direct explanation for the low oscillation strength of the interlayer peaks: due to the small spilling of the Mo (W) states into the valence (conduction) band edge, the corresponding optical matrix element is forced to be small compared to the intralayer transitions, where the spatial overlap of the involved states is significantly larger.

The responses of these transitions to polarized light can be understood from symmetry considerations. For all three stacking orders, the point group of the bands at the $K$-point is $C_{3v}$, which is also the point group of the MoSe$_2$/WSe$_2$ for all three studied stacking orders. According to the corresponding character table, the irreducible representations of Mo $d_{x^2-y^2}$ and $d_{xz}$ is $E$, the representation of Mo $d_{z^2}$ orbitals is $A_1$ and the x- and z-axes transform as $E$ and $A_1$, respectively.\FloatBarrier

As a general rule, an optical transition for a given light polarization is allowed by symmetry, if the product of the irreducible representations of the polarization direction and the initial and final states in turn contains the totally symmetric irreducible representation, in this case $A_1$.\\

We then obtain for the strongest contribution of AA stacking, see Fig.~\ref{fig:exc-cont}:
\begin{itemize}
\item $d_{x^2-y^2}\rightarrow d_{z^2}$ with in-plane polarized light: \begin{equation*} E\otimes E \otimes A_1=A_1+A_2+E, \end{equation*} \emph{i.e.} symmetry allowed.
\item $d_{x^2-y^2}\rightarrow d_{z^2}$ with out-of-plane polarized light: \begin{equation*} E\otimes A_1 \otimes A_1=E, \end{equation*} \emph{i.e.} symmetry forbidden.
\end{itemize}

For AA' stacking:
\begin{itemize}
\item $d_{xz}\rightarrow d_{z^2}$ with in-plane polarized light: \begin{equation*} E\otimes E \otimes A_1=A_1+A_2+E, \end{equation*} \emph{i.e.} symmetry allowed.
\\
\item $d_{xz}\rightarrow d_{z^2}$ with out-of-plane polarized light: \begin{equation*} E\otimes A_1 \otimes A_1=E, \end{equation*} \emph{i.e.} symmetry forbidden.
\end{itemize}

And for AB stacking:
\begin{itemize}
\item $d_{z^2}\rightarrow d_{z^2}$ with in-plane polarized light: \begin{equation*} A_1\otimes E \otimes A_1=E, \end{equation*} \emph{i.e.} symmetry forbidden.
\item $d_{z^2}\rightarrow d_{z^2}$ with out-of-plane polarized light: \begin{equation*} A_1\otimes A_1 \otimes A_1=A_1, \end{equation*} \emph{i.e.} symmetry allowed.
\end{itemize}

Similar considerations hold for the other contributions, for example the $Se^{Mo}_x \rightarrow Se^{Mo}_x$ for AA stacking.\\

The origin of the optical strength of interlayer excitons in small fractions of stacking-dependent interlayer hybridization further explains the difference in polarization behavior between interlayer excitons and the intralayer excitons, even though the involved bands are the same: for the intralayer excitons, the oscillation strength arises from transitions between valence and conduction bands that belong to the same layer and whose composition, apart from a small hybridization with the neighbouring layer, is almost completely independent of the stacking order (see Tab.~\ref{tab:contrib}). The oscillation strength and response to light polarization hence is largely independent of the stacking order as well.

\section{M{\MakeLowercase o}S$_2$-WS{\MakeLowercase e}$_2$ heterostructure}
\begin{figure*}
\centering
\includegraphics*[width=0.44\textwidth]{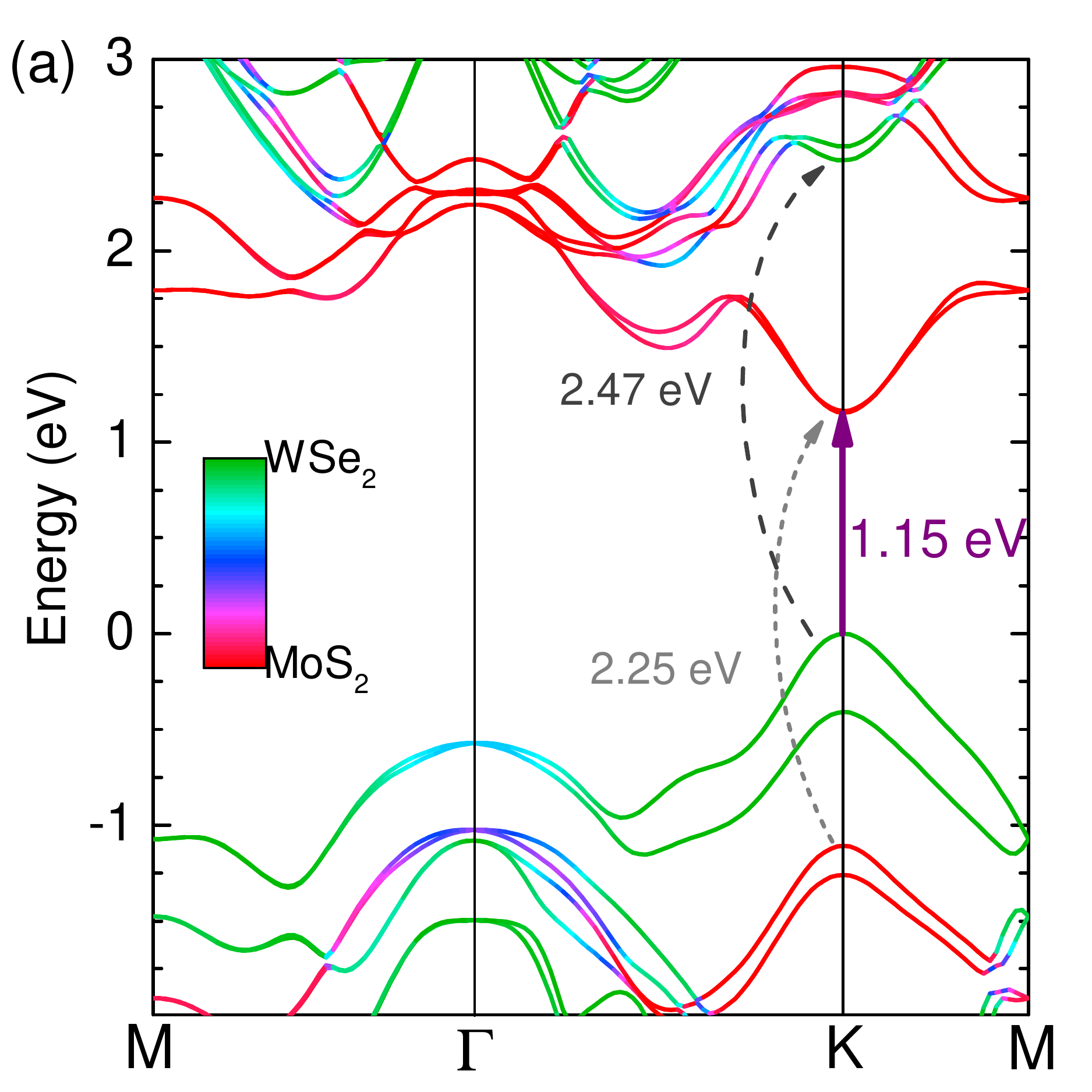}
\includegraphics*[width=0.55\textwidth]{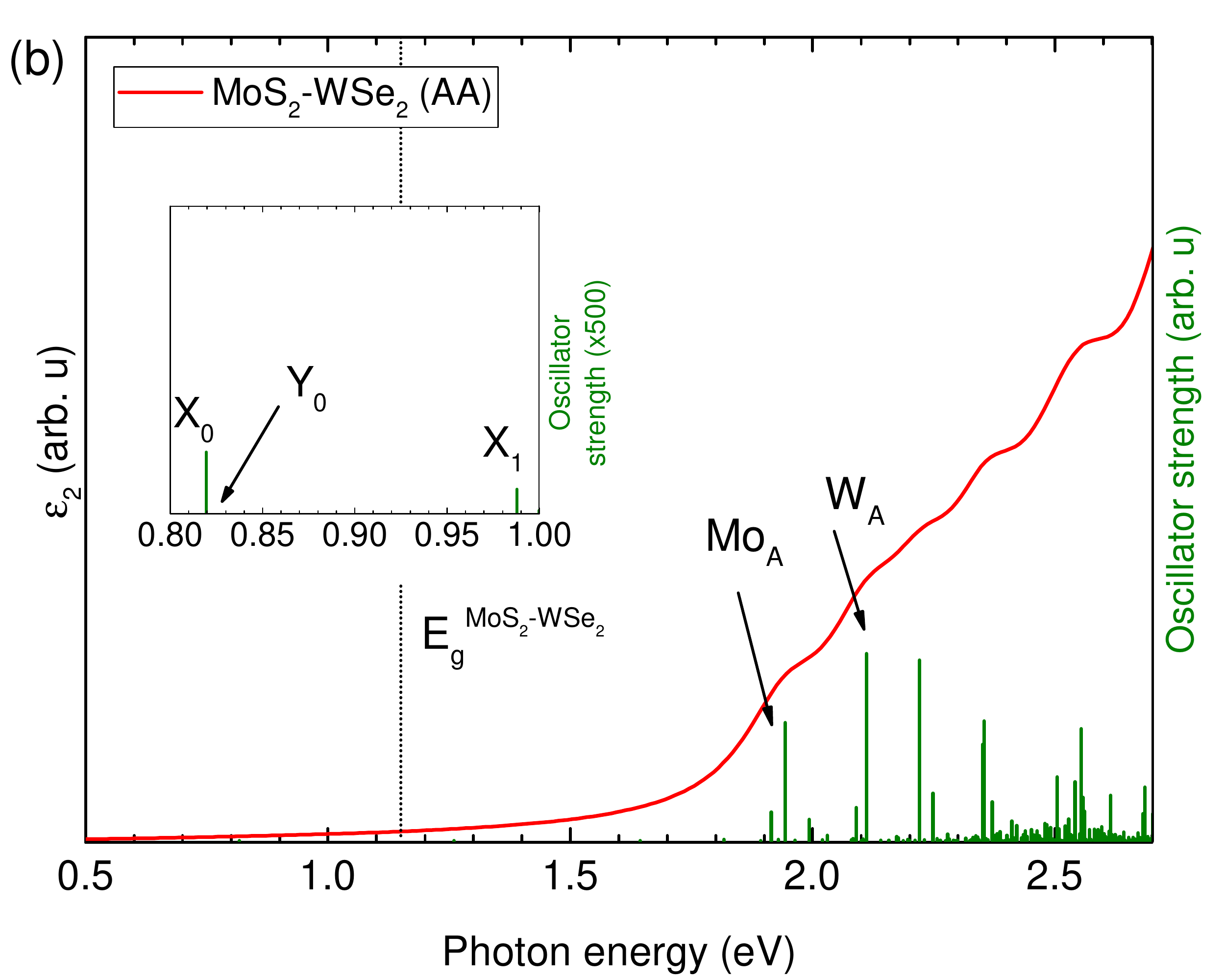}
\caption{\label{fig:MoS2-WSe2-bands_absorption} (Color online) (a) Electronic bandstructure of an AA-stacked MoS$_2$/WSe$_2$ heterostructure as calculated using the G$_0$W$_0$ approximation and spin-orbit interaction. The relative contributions of the two materials to the electronic bands are shown through a colour code, where red corresponds to 100\% MoS$_2$ and green to 100\% WSe$_2$. (b) Corresponding dielectric function including electron hole-effects and spin-orbit interaction. As for the MoSe$_2$/WSe$_2$ heterostructures, we used a 21x21 k-point grid for the solution of the Bethe-Salpeter Equation. The inset shows the energy region around the zero-order interlayer excitons, X$_0$ and Y$_0$.}
\end{figure*}
Besides the optical spectra MoSe$_2$/WSe$_2$ heterostructures in the main manuscript, we also decided to do simulations on stacked MoS$_2$/WSe$_2$ heterostructures. An obvious hurdle is the significant lattice mismatch between monolayer MoS$_2$ (in-plane lattice constant a=3.16\,\AA in our calculations) and monolayer WSe$_2$ (a=3.926\,AA) that requires large supercells for a strain-free geometry. In order to keep the calculations feasible and obtain at least qualitative results, we decided to use a similar approach as in the main text and produce and AA-stacked structure from a unit cell of MoS$_2$ and a unit cell of WSe$_2$. Structural relaxation produced and intermediate lattice constant of 3.22\,\AA and a distance between the Mo and the W layers of 6.88\,AA. Correspondingly, the MoS$_2$ is subject to about 2\% of tensile strain, while the WSe$_2$ layer is compressed by 2\%.

Figure~\ref{fig:MoS2-WSe2-bands_absorption}~(a) shows the calculated electronic bandstructure of the heterostructure from the G$_0$W$_0$ approximation. In contrast to the MoSe$_2$/WSe$_2$ heterostructure, the MoS$_2$ and WSe$_2$ form a type-II heterojunction with a direct fundamental band gap of 1.15\,eV, with both valence band maximum and conduction band minimum are at the $K$ point in the hexagonal Brillouin zone. We note that due to the imposed strain, we expect the intralayer band gap in the MoS$_2$ layer to be reduced compared to the strain-free structure, while the intralayer band gap in the WSe$_2$ layer should be overestimated. This can be seen from comparison of our obtained MoS$_2$ and WSe$_2$ band gaps of 2.25\,eV and 2.47\,eV, respectively, with values of the AA-stacked MoSe$_2$/WSe$_2$ heterostructure in the main text (WSe$_2$: 2.011\,eV, MoSe$_2$: 2.015\,eV). It is thus likely that the value of the interlayer band gap is underestimated compared to a strain-free layer alignment. Due to the larger difference in band energies compared to the vacuum level, the valence and conduction band offsets are significantly larger than in the MoSe$_2$/WSe$_2$ heterostructures and the WSe$_2$ and MoS$_2$ valence bands at the $K$ point are cleanly separated. This agrees well with the recent report by Lantini \emph{et al.}~\citep{lantini-2017_suppl}. 
\begin{figure*}
\centering
\includegraphics*[width=0.95\textwidth]{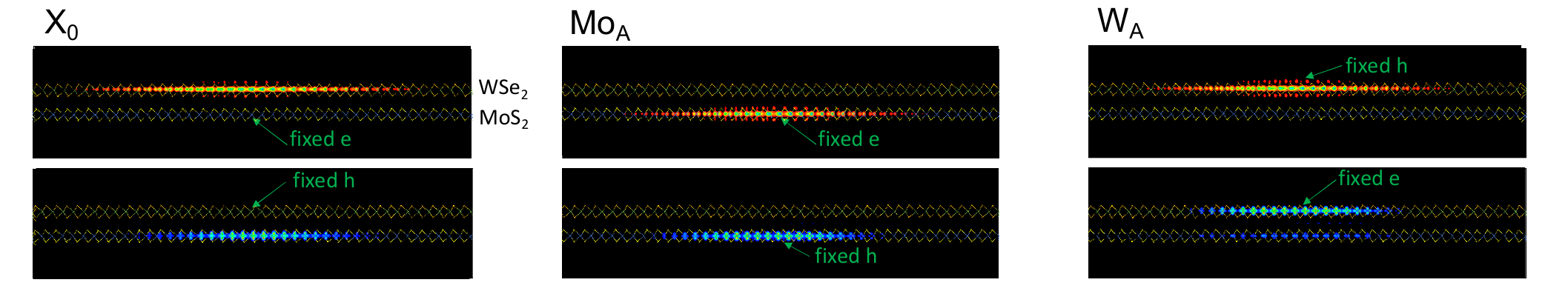}
\caption{\label{fig:MoS2-WSe2-wavefunctions} (Color online) Electron (blue) and hole (red) contributions to the excitonic wave functions of the \textsl{X$_0$}, \textsl{Mo$_A$}, and \textsl{W$_A$} excitons of an AA-stacked MoS$_2$/WSe$_2$ heterojunction. The excitonic wave functions were computed for a supercell of 21x21 unit cells and projected onto the x-z plane. For the electron (e) [hole (h)] contributions, the hole [electron] was fixed at the Mo  [W] atom indicated by the green arrows. }
\end{figure*}

Using the obtained electronic structures, we proceeded to calculate the dielectric functions including electron-hole effects and spin-orbit coupling. Figure~\ref{fig:MoS2-WSe2-bands_absorption}~(b) shows the obtained spectrum with a decomposition into the contributing peaks. The onset of strong absorption is dominated by two bright transitions, which we identify as the \textsl{A} exciton of the MoS$_2$ layer (\textsl{Mo$_A$}) and the \textsl{A} exciton of the MoS$_2$ layer (\textsl{Mo$_A$}). Due to the imposed strain, the order of the MoSe$_2$ and WSe$_2$ peaks are reversed compared to experimental observations, even if spin-orbit coupling is included. Like for the MoSe$_2$/WSe$_2$ heterostructure, we find two Rydberg series \textsl{X$_n$} and \textsl{Y$_n$} of excitonic transitions below the MoSe$_2$ and WSe$_2$ \textsl{A} excitons. The energetically lowest of the transitions, \textsl{X$_0$}, appears at an energy of 0.82\,eV. Analysis of the excitonic wavefunctions, Fig.~\ref{fig:MoS2-WSe2-wavefunctions}, show the same charge separation of electron and hole contributions as we found for the corresponding transitions in the MoSe$_2$/WSe$_2$ heterostructure, while the wavefunctions of \textsl{Mo$_A$} and \textsl{W$_A$} are confined to the MoSe$_2$ and WSe$_2$ layers, respectively. 
We again neglected spin-orbit interaction in order to estimate the exciton binding energies on a denser 33x33 k-point grid. The derived binding energy of the \textsl{X$_0$} of $E_b^{X_0}$=310\,eV is slightly larger than the binding energy in the MoSe$_2$/WSe$_2$ heterostructure and the results by Lantini \emph{et al.} as obtained from a modified Wannier model based on electron and hole effective masses and dielectric screening obtained from G$_0$W$_0$ calculations~\citep{lantini-2017_suppl}. Our binding energies for \textbf{$Mo_A$} and \textsl{W$_A$} (on the order of 310\,meV) are almost the same as that for $E_b^{X_0}$, smaller by about 0.1,\,eV than those from Ref.~\onlinecite{lantini-2017_suppl}. The difference might be related to the relative rotation of the layers and strain-free geometry in Ref.~\onlinecite{lantini-2017_suppl} in contrast to our strained AA-stacked configuration. On the other hand, our approach offers a superior description of electron-hole interaction effects and the qualitative findings are very similar to the (strain-free) MoSe$_2$/WSe$_2$ heterostructures, implying some merit in our results. We thus attribute the strongly underestimation of the \textsl{X$_0$} peak position compared to the \textsl{I} peaks from experiments~\citep{MoS2-WSe2-1_suppl, MoS2-WSe2-2_suppl} at 1.4-1.5\,eV to an underestimation of the interlayer band gap. 

%



\begin{thebibliography}{23}
\bibitem [{NoteS2()}]{NoteS2}%
  \BibitemOpen
  \bibinfo {note} {The calculations used a 12x12 k-point grid, one valence and
  one conduction band, and included local field effects up to an energy of 100
  \protect \tmspace +\thinmuskip {.1667em}eV. The conduction band energies were
  corrected by a scissor shift of 1.0\protect \tmspace +\thinmuskip
  {.1667em}eV.}\BibitemShut {Stop}%
\end{thebibliography}
\end{document}